\documentclass{article}
\usepackage{amsmath,amssymb}
\usepackage{graphicx}
\usepackage{color}
\usepackage{wrapfig}
\usepackage{url}

\usepackage{lineno}

\usepackage{array} 
\usepackage{float}
\usepackage{authblk}
\usepackage{multirow}
\usepackage[symbol]{footmisc}

\title{Forecasting racial dynamics at the neighborhood scale using Density-functional Fluctuation Theory}

\author[1]{Yunus A. Kinkhabwala}
\author[2]{Boris Barron}
\author[3,4]{Matthew Hall}
\author[2]{Tomas A. Arias}
\author[2]{Itai Cohen}
\affil[2]{\footnotesize Department of Physics, Cornell University, Ithaca, NY 14853, USA.}
\affil[1]{\footnotesize Department of Applied and Engineering Physics, Cornell University, Ithaca, NY, 14853, USA.}
\affil[3]{\footnotesize Policy Analysis and Management, Cornell University, Ithaca, NY, 14853, USA.}
\affil[4]{\footnotesize Cornell Institute for Public Affairs, Cornell University, Ithaca, NY, 14853, USA.}

\begin{document}

\maketitle

\newcommand{\com}[1]{\textcolor{red}{(#1)}}
\newcommand{\comc}[2]{\textcolor{#1}{(#2)}}

\newpage

\section{Abstract}


Racial residential segregation is a defining and enduring feature of U.S. society, shaping inter-group relations, racial disparities in income and health, and access to high-quality public goods and services \cite{wellman2014transportation,anderson2003providing,oka2014capturing,kershaw2011metropolitan,massey1993american,krysan2017cycle}. 
The design of policies aimed at addressing these inequities would be better informed by descriptive models of segregation that are able to predict neighborhood scale racial sorting dynamics \cite{hauer2016millions,wilson2016evaluation}.
While coarse regional population projections are widely accessible \cite{smith2013practitioner,wachter2014essential,raftery2012bayesian, crone2011advances}, small area population changes remain challenging to predict because granular data on migration is limited and mobility behaviors are driven by complex social and idiosyncratic dynamics \cite{vasan2018use,chi2017small,wilson2016evaluation}. Consequently, to account for such drivers, it is necessary to develop methods that can extract effective descriptions of their impacts on population dynamics based solely on statistical analysis of available data. Here, we develop and validate a Density-Functional Fluctuation Theory (DFFT)\cite{mendez2018density,ChenArxiv} that quantifies segregation using density-dependent functions extracted from population counts and uses these functions to accurately forecast how the racial/ethnic compositions of neighborhoods across the US are likely to change. Importantly, DFFT makes minimal assumptions about the nature of the underlying causes of segregation and is designed to quantify segregation for neighborhoods with different total populations in regions with different compositions. This quantification can be used to accurately forecast both average changes in neighborhood compositions and the likelihood of more drastic changes such as those associated with gentrification and neighborhood tipping\cite{ellen2016cityscape,freeman2009neighbourhood,crowder2000racial}. As such, DFFT provides a powerful framework for researchers and policy makers alike to better quantify and forecast neighborhood-scale segregation and its associated dynamics.

\section{Introduction}
Various methods have been developed to accurately forecast demographic changes at regional scales \cite{poot2016gravity,smith2013practitioner,wachter2014essential,raftery2012bayesian,crone2011advances,hauer2019population}, but determining how these changes trickle down to the neighborhood scale where the effects of segregation are most apparent and pernicious remains a major challenge \cite{vasan2018use,chi2017small,wilson2016evaluation}. 
This challenge is due to limitations in access to sufficient data as well as difficulties in modeling the decision making process of residential mobility. If complete counts of individual migration events were available, the migration rates that predominantly drive neighborhood-scale population changes could be directly measured and population dynamics could be simulated. Due to data and privacy limitations that conceal migration events, however, such an approach using publicly available data is not possible. Alternatively, if the drivers of human mobility were fully understood, one could reliably model changes in neighborhood populations. Often, however, the drivers of residential choice are complex and multifaceted, ranging from housing policies to social preferences and economic constraints, rendering such approaches unreliable. In the absence of reliable models, researchers have resorted to using a variety of \emph{ad hoc} approaches to generate small area population projections, ranging from estimations of birth, death, and migration rates\cite{wilson2016evaluation}, to extrapolations\cite{grauwin2012dynamic,armstrong2001extrapolation}, population share models \cite{wilson2015new}, microsimulations\cite{tanton2014review}, regressions\cite{chi2017small}, and GIS models\cite{baker2014spatial,vasan2018use}. These methods, however, typically do not account for segregation, which is known to drive residential dynamics. Whether it is possible to develop accurate models of population dynamics driven by segregation from available coarse grained data of neighborhood scale population counts remains an open question.  

An important step towards developing such a model entails determining how much information relating to segregation can be extracted from available data sets. While there is a rich and detailed history of methods aimed at quantifying neighborhood-scale segregation and understanding its causes, these methods typically reduce segregation trends to simple scalar indices\cite{james1985measures,massey1988,reardon2002measures,reardon2004measures,yao2019spatial,s2016segregation,hennerdal2017multiscalar,chodrow2017structure,roberto2015divergence}. Impressively, such indices have provided important tools for understanding how segregation differs across cities, how it has changed over time, how it correlates with social and economic outcomes, and how drivers of segregation are likely to change into the future\cite{ellis2018predicting}. However, using a single index to describe population distributions is problematic since a single index cannot distinguish between segregation scenarios that can lead to drastically different changes in neighborhood populations (See supplementary figure 1). For example, consider two hypothetical cities with a population evenly split between two subgroups, red and blue, living in equally-sized neighborhoods. One city is comprised of equal numbers of neighborhoods whose residents are entirely blue, entirely red, and highly mixed (50/50); while the other city is comprised of equal numbers of neighborhoods whose residents are 83\% red and 83\% blue. These two cities clearly have different segregation patterns between the red and blue population subgroups and yet are represented by the exact same dissimilarity index. Moreover, migration patterns of these two cities must differ in order to sustain these population distributions as individuals move between neighborhoods. In the first city, a neighborhood that is populated entirely by blue persons would likely remain entirely blue while the same such neighborhood in the second city, in which there are typically no such neighborhoods, would evolve towards the more probable composition closer to 83\% blue. Segregation indices, then, are overly reductive in trying to capture such differences in neighborhood migration dynamics. The solution to this problem is to move beyond simple segregation indices and develop more detailed measures, such as segregation functions, that can distinguish between such scenarios. These more detailed measures could then be used to develop methods to predict population dynamics at the neighborhood scale. 

\begin{figure}
    \includegraphics[width=6.0in,keepaspectratio]{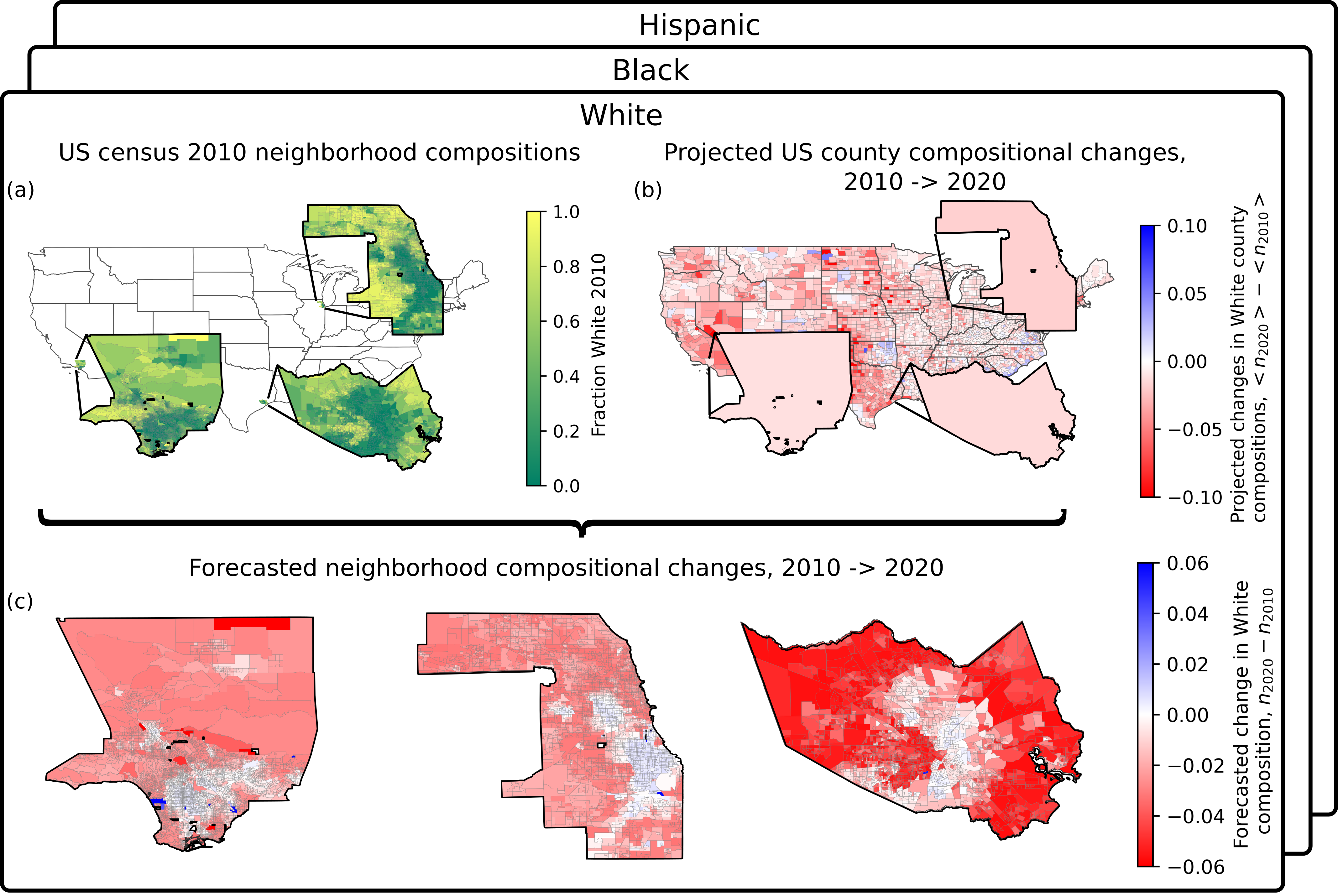}
    \caption{\textbf{Data-driven forecasting of changes in neighborhood racial/ethnic compositions.} 
    (\textbf{a}) Neighborhood-scale (US census block group) White compositions in 2010 for the three most populous counties, Los Angeles, Cook, and Harris. (\textbf{b}) Projected changes in county-scale White compositions from 2010 to 2020 using traditional demographic methods\cite{hauer2019population}. (\textbf{c}) Forecasts of changes in neighborhood composition for counties shown in (a). Here, we infer segregation functions from data in (a) that, when incorporated with traditional projections of coarser county-scale changes in composition (b), can generate accurate forecasts of changes in neighborhood-scale racial compositions. We perform this analysis for the White, Black, and Hispanic population subgroups and verify the validity of such forecasts by projecting 2010 compositions based on data from 2000. 
    }
    \label{fig:1}
\end{figure}

Here, we show such an approach using available data to capture the effects of segregation on small area population changes is possible. We do so by developing a statistical framework based on Density-Functional Fluctuation Theory (DFFT) \cite{ChenArxiv} that extracts functions describing segregation from available demographic data and uses these functions to accurately predict neighborhood scale population changes (Fig.~\ref{fig:1}). Specifically, we use observations of fluctuations in local neighborhood compositions (Fig.~\ref{fig:1}a) to extract functions that quantify neighborhood-scale interactions between racial/ethnic subgroups. We then use projected changes in county compositions (Fig.~\ref{fig:1}b) in combination with the inferred segregation functions to forecast how neighborhoods within these regions are likely to change their compositions. As a proof of concept, we validate this approach using historical census data and then go on to make predictions of likely changes in racial/ethnic compositions at the neighborhood scale across the entire United States for the year 2020 (Fig.~\ref{fig:1}c) and beyond.   

\section{Results}
The approach we pursue here entails estimating the probabilities of neighborhood racial/ethnic compositions underlying census population counts and then determining if such probabilities can be used to predict the tendencies of neighborhoods to change their racial composition. For a given county, such a probability distribution can be expressed as $\sim e^{-H(n,s)}$ where $H$ is a function that penalizes the composition (fraction) of persons identifying with a given subgroup, $n$, and that depends on the total number of people in a neighborhood, $s$. Since there are a limited number of neighborhoods within a given county, each with its own total population, we can improve the statistical sampling of the data by separating from $H$ the dependence on $s$. One known statistical effect associated with different population counts is that certain compositions are more likely than others based on the number of distinct arrangements of individuals within a neighborhood that give rise to that composition. This effect can be accounted for using a binomial factor, $\binom{s}{sn}$, which counts the number of unique arrangements leading to a given composition. Second, if we make a mean-field approximation that $H(n)$ is the penalty for any given person to be found in a neighborhood with composition $n$, then the joint penalty for simultaneously observing $s$ individuals in that neighborhood is simply $sH(n)$. Thus, we can approximate the probability $P(n;s)$, of observing a composition, $n$, in a neighborhood with $s$ total persons as  
\begin{equation}
    P(n;s) = z^{-1}\binom{s}{sn}e^{-sH(n)},\label{eq:probability_distribution}
\end{equation}
where $z$ is a normalization constant \cite{ChenArxiv}. 

\begin{figure}
    \includegraphics[width=5.5in,keepaspectratio]{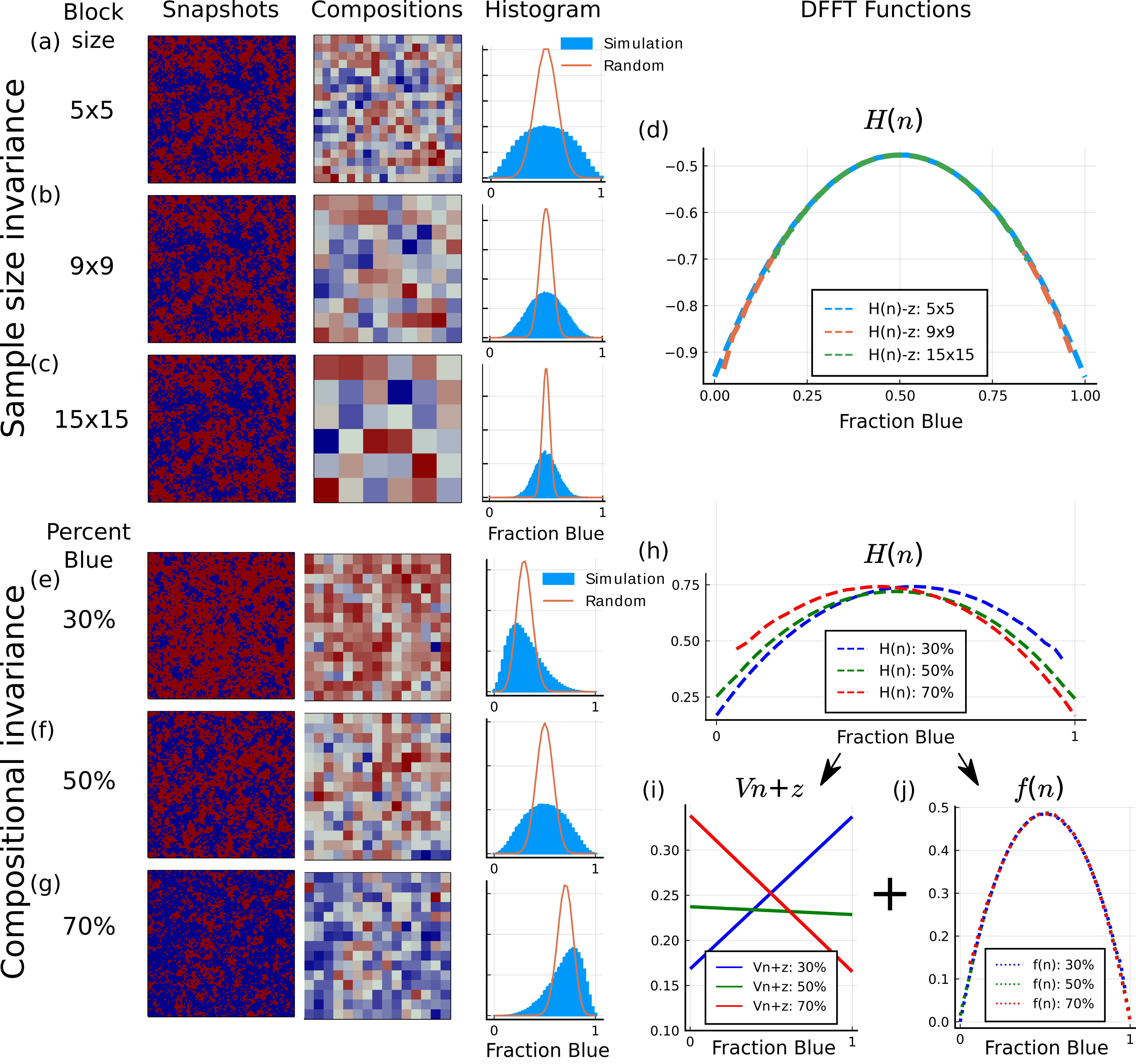}
    \caption{\textbf{Demonstration of (A) sample size invariance and (B) compositional invariance of DFFT functions using simulated data.} 
    (\textbf{Top half}) Schelling simulations of segregated agents (Snapshots) grouped into different neighborhood/sample sizes (Compositions) with 25 (\textbf{a}), 81 (\textbf{b}), and 225 (\textbf{c}) agents. Probability distributions (Histograms) tallied using 1000 snapshots (blue bars) and distributions expected for random data (red curves). The deviations of these probability distributions from random capture segregation tendencies but distributions appear to vary by neighborhood size. (\textbf{d}) DFFT functions for each neighborhood size, $H(n)$, obtained through an appropriate transformation of the probability distributions. Overlap of these functions demonstrates sample-size invariance. (\textbf{Bottom half}) Simulated "counties" (Snapshots) with identical agent interactions but varying county-scale compositions; 30\% (\textbf{e}), 50\% (\textbf{f}), and 70\% (\textbf{g}) blue agents. (Compositions) Neighborhood compositions grouped using a neighborhood size of 25 agents. (Histograms) Probability distributions (blue bars) tallied using 1000 snapshots and the distribution expected for random data (red curves). (\textbf{h}) DFFT functions, $H(n)$, obtained through an appropriate transformation of the histograms vary only by a linear term, $Vn+z$ (\textbf{i}), while the higher order terms that capture the segregation interactions $f(n)$, are compositionally invariant as shown by the overlap of these functions (\textbf{j}).}
    \label{fig:2}
\end{figure}

To investigate this neighborhood-size invariant \emph{ansatz} for the function $H(n)$, we test whether a single penalty function can account for simulated data of systems with different neighborhood sizes (Fig.~\ref{fig:2}). Specifically, we conduct a Schelling model simulation of a segregated county \cite{ChenArxiv} (See Methods for details), divide the county into equal sized neighborhoods, and count the number of agents in each neighborhood. Finally, we generate a histogram of the number of times each neighborhood composition is observed (blue), and extract the penalty function $H(n)$ through simple rearrangement of equation \ref{eq:probability_distribution}. This procedure is repeated for the same simulation, but with different neighborhood sizes (5x5, 9x9, and 15x15). We find that, regardless of the neighborhood size, we extract the same penalty function, $H(n)$. Importantly, the invariance of this statistical measure to the number of persons in a neighborhood, known in the literature as sample size invariance\cite{reardon2002measures}, allows for efficiently inferring the probability distribution underlying the Census population counts for a given county regardless of the neighborhood size $s$.

Importantly, while the penalty function $H(n)$ provides an effective description of the emergent interactions between racial/ethnic subgroups, it is susceptible to changes in the county composition. For example, two counties with different racial/ethnic compositions may have the same neighborhood-scale interactions between racial/ethnic subgroups, but drastically different probability distributions simply because there are proportionally more residents of a particular subgroup in one county than the other. To address this issue, we make an additional assumption that the penalty function can be broken up into a term linear in $n$ that fully accounts for variations in the proportions of the different subgroups in the county, whose slope we call the "vexation, $v$," and an interaction term that quantifies the tendencies of neighborhoods to adopt a particular composition, which we call the "frustration function" $f(n)$, 
\begin{equation}
    H(n) = vn+f(n)\label{eq:headache_decomposition}.
\end{equation}
Here, the first term acts as a constant penalty for increases in the composition and so counties with higher values for the vexation must have lower overall compositions of the chosen subgroup. The frustration function, $f(n)$, contains the higher order terms of $H(n)$ and acts as a penalty function that captures segregation tendencies of neighborhoods\cite{mendez2018density, ChenArxiv}.
In particular, when $f(n)=0$, the distribution is a random binomial distribution. When $f(n)$ has negative curvature, it drives neighborhoods towards segregated compositions resulting in broader probability distributions. To demonstrate the utility of this separation, we simulate three counties as before, but this time with the same neighborhood size and segregation behaviors but different overall compositions (Fig.~\ref{fig:2}e-g) and calculate the corresponding $H(n)$ functions (Fig.~\ref{fig:2}h). While the $H(n)$ functions are distinct, subtraction by linear functions with slope $v$ shows they can all be described by the same curvature functions, $f(n)$, that quantify neighborhood scale segregation behaviors. The independence of this $f(n)$ function to changes in county composition, referred to generally as compositional invariance \cite{reardon2002measures}, allows for 1) combining data from counties with different overall compositions under the assumption that the neighborhoods within them share the same segregation tendencies, and 2) predicting how the probability distribution for a given county should change if the overall composition of the county changes but its segregation tendencies remain the same. 

\begin{figure}
    \centering
    \includegraphics[width = 6in]{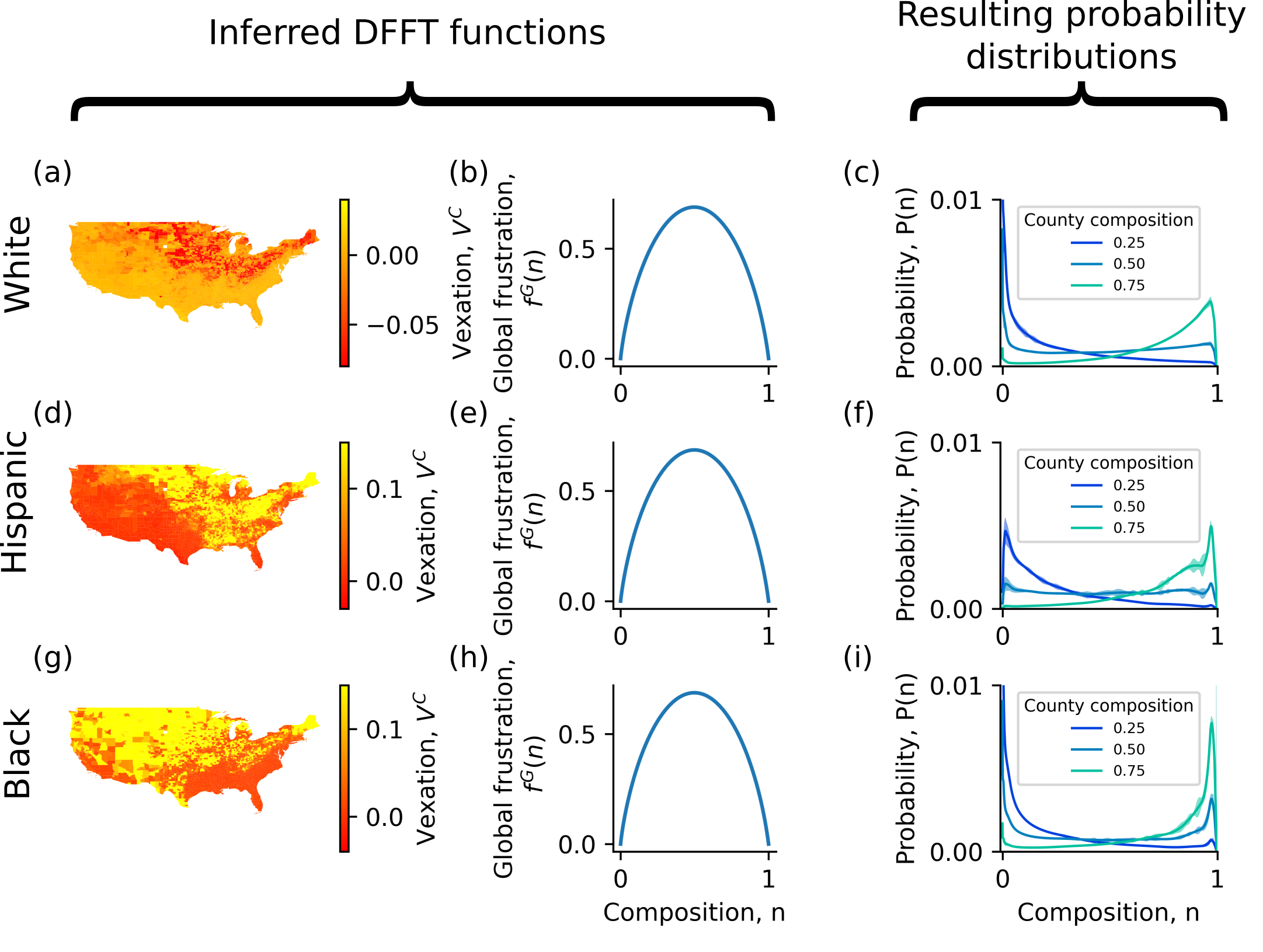}
    \caption{\textbf{DFFT functions applied to US census data quantify tendencies of racial/ethnic subgroups towards different counties and tendencies to segregate.} 
    (\textbf{a,d,g}) Vexations, $v^{(C)}$, of the Black, Hispanic, and White subgroups respectively for the contiguous US. Counties with lower vexations have higher compositions of the respective subgroup. (\textbf{b,e,h}) Frustration functions, $f^{(G)}(n)$, for the three respective subgroups, all of which possess negative curvature indicative of segregation. Uncertainties of frustration functions are calculated using bootstrapping (Methods) but are too small to see. (\textbf{c,f,i}) Hypothetical probability distributions for 1000 person neighborhoods with county compositions of 25\%, 50\% or 75\%. Distributions are calculated using the respective frustration functions (b,e,h) and vexation values that match the desired county composition. The spread of these distributions toward segregated values provides an interpretation of the DFFT segregation functions. Uncertainties propagated from the frustration function are shown for all probability distributions but for most compositions are on the order of the thickness of the line. See supplemental figures for clearer visualization of uncertainties. 
    }
    \label{fig:3}
\end{figure}

Importantly, the size and composition invariant properties of this DFFT framework allow for easily consolidating data and making comparisons between different neighborhood compositions. Specifically, application of the DFFT form to US census population counts leads to both vexation values (Fig.~\ref{fig:3} left column) representing county compositions of racial subgroups and a frustration function (Fig.~\ref{fig:3} middle column) characterizing segregation tendencies of a chosen subgroup at the neighborhood scale. The data we use are Census block group population counts (typically between 500-2,500 total individuals) categorized according to the three largest racial/ethnic subgroups, Hispanics and Non-Hispanic Whites, Blacks, and Others. The Census block group is the smallest spatial scale for which longitudinal changes can be observed due to limitations in the publicly available data.  To consolidate data from neighborhoods with different total populations and within different counties, we use a maximum likelihood approach (Methods) that extracts DFFT functions from population counts for each subgroup separately. As expected, counties with high compositions of a given subgroup have low vexation values and vice versa. Regionally, the vexation of the White, Hispanic and Black subgroups are least in the Midwest, West, and South respectively. We simultaneously extract frustration functions for each subgroup and find that they all have negative curvatures, which indicates significant neighborhood scale segregation between each subgroup and the remainder of the population. For this analysis, we assume that all counties share the same "global" frustration function, $f^{(G)}(n)$ for a given racial/ethnic subgroup, thus defining a measure of the "average" segregation across the nation. This shared frustration function allows for the inclusion of neighborhoods from all counties across the entire nation, including those counties with small populations or dominated by one racial group for which the domain of possible compositions is not well represented. We find through analysis of random subsets of counties (see Computational Methods) that, despite differences in segregation across the US, this global frustration captures the dominant trends for these population subgroups. Using the above inferred DFFT functions, it is now possible to generate the probability of observing neighborhood compositions for a county with any hypothetical composition. Specifically, for each global frustration function we tune the vexation value to generate probability distributions for 1000 person neighborhoods in hypothetical counties with compositions of 25\%, 50\%, or 75\% (Fig.~\ref{fig:3} right column). Distributions for counties with 50\% composition are spread out towards segregated compositions while distributions for counties with 25\% and 75\% composition are peaked at compositions near zero and one respectively. As such, these distributions provide an intuitive representation of segregation. Differences between these distributions across the racial/ethnic subgroups, shown in more detail in the supplemental materials, further demonstrate that small differences between the global frustration functions give rise to significantly different distributions. Importantly, these distributions and the differences between them capture trends previously reported for these racial/ethnic subgroups. For example, we find that Black populations experience the greatest amount of segregation while Hispanic populations experience the least amount of segregation \cite{parisi2011multi} (for additional examples see Supplementary materials). Importantly, such hypothetical distributions reflect an average segregation behavior across the United States and indicate whether an actual neighborhood with a specific composition, $n$, is probable or improbable. 

This quantification of neighborhood-scale demographics presents a remarkable opportunity for making forecasts of future compositions. In particular, assuming the global frustration function changes slowly over time, we expect that neighborhoods will tend towards more likely compositions but with significant fluctuations. Specifically, given the penalty function $H(n)$ associated with a particular hypothetical distribution, its derivative, $H'(n)$, quantifies the tendency of a neighborhood with a given composition to undergo a small change in $n$. As such, we can use a Metropolis-Hastings algorithm to formulate a set of recursive equations for evolving a neighborhood's probability distribution where, at each step, the neighborhood composition $n$ transitions to lower $n - \delta n$ or higher $n + \delta n$ values with the following probabilities $p(n)$: 

\begin{align}
    p(n \rightarrow n-\delta n) &= \frac{n}{1+e^{-H'(n)}} \label{eq:rates1}\\
    p(n \rightarrow n+\delta n) &= \frac{1-n}{1+e^{H'(n)}} \label{eq:rates2},
\end{align}

\begin{figure}
    \centering
    \includegraphics[width = 6.5 in]{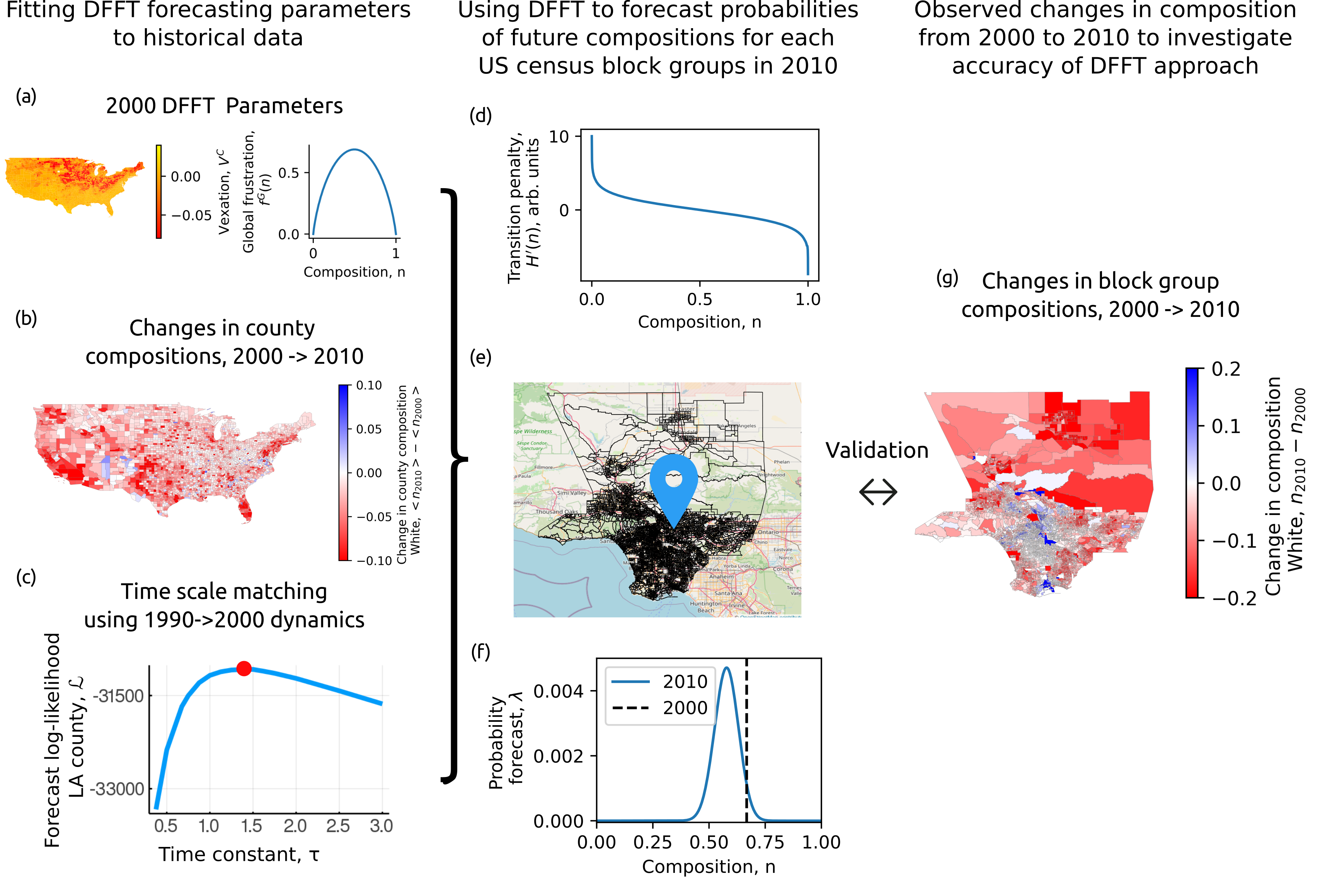}
    \caption{\textbf{Analysis pipeline using DFFT functions to forecast changes in block group compositions.}
    (\textbf{a}) DFFT functions extracted from year 2000 population counts for White/non-White population. (\textbf{b}) Changes in the overall composition of the counties from the years 2000 to 2010. (\textbf{c}). Log-likelihood metric of forecast accuracy of Cook county 2000 compositions using 1990 DFFT functions plotted against the time scale, $\tau^{(C)}$, with optimum time scale (red point) for Cook county. (\textbf{d}) Transition penalty is the derivative of the $H(n)$ function built using parameters from (\textbf{a-c}). High values for $H'(n)$ indicate a low probability for increases in composition. (\textbf{e}) Randomly chosen neighborhood in Cook county (map marker). (\textbf{f}) Forecast of composition change of chosen neighborhood. Given the initial composition in 2000 (vertical black dotted line), DFFT forecasts the probability of observing any composition in the year 2010 (blue curve). (\textbf{g}) Our historical forecasts can be validated by means of comparisons to observed changes in neighborhood composition from 2000 to 2010, which is the subject of Fig.~5.}
    \label{fig:4}
\end{figure}

Intuitively, the numerator in Eq.~\ref{eq:rates1} reflects the composition of the chosen subgroup while the numerator in Eq.~\ref{eq:rates2} reflects the remaining subgroups. Thus, in a predominantly White neighborhood it becomes more likely for a White person to be replaced by a non-White person simply because there are more White people than non-White people. The denominators, on the other hand, capture the segregation tendencies of neighborhoods independently of such trivial statistical effects. For example, if $H'(n)$ is large and positive, then the probability of increasing the composition $n$ becomes small (Eq.~\ref{eq:rates1}). These equations obey detailed balance which guarantees that under reasonable conditions (see methods) any neighborhood, regardless of its initial composition, will evolve its probability distribution from being peaked at the initial composition to the distribution defined by $H(n)$. More interestingly, over a smaller number of iterations, these equations can forecast which compositions are likely to undergo large changes and the relative rates at which these changes will occur. 
This framework can make such forecasts while also accounting for shifts in county compositions in a natural way. Since $H'$ depends on the vexation, changes in county composition (Fig.~\ref{fig:4}b) are incorporated via an additional constant shift to the vexation for each county, $v^C \rightarrow v^C + \mu^C$, that constrains the forecast to match expected shifts in county demographics. For example, if the county composition is expected to increase then adding a constant $\mu < 0$ to the vexation will uniformly shift $H'$ down making it more likely for any neighborhood, regardless of its composition, to increase $n$. Importantly, these constant shifts do not affect the tendency of neighborhoods to segregate as defined by the frustration function $f(n)$. Further, these $\mu$ values are constrained to match predefined changes in county composition over a given number of iterations of Eqs.~\ref{eq:rates1},\ref{eq:rates2}, such as those predicted by traditional demographic projections at the county scale (Fig.~\ref{fig:4}b)\cite{hauer2019population}. In this manner, the DFFT framework is able to bridge scales from readily available demographic projections at the county scale to compositional changes driven by segregation at the neighborhood scale.

As a final step towards achieving neighborhood-scale forecasts, we calibrate the timescale associated with evolving the neighborhood-scale probability to the annual time scale extracted from observing previous changes in compositions. In figure~\ref{fig:4}c we plot the log-likelihood metric of forecast accuracy (Methods) for neighborhood changes observed from 1990 to 2000 as a function of the time constant, $\tau^{(C)}$, for a given county $C$ that scales the number of iterations using equations~\ref{eq:rates1},~\ref{eq:rates2} needed for a given neighborhood to change its composition in a year. To account for differences in mobility among counties, we choose the value of $\tau^{(C)}$ that optimizes the forecast accuracy independently for each county. Assuming these time constants do not change significantly on the decadal scale, this calibration can be used to make forecasts over subsequent decades. 

Using this framework, we forecast neighborhood composition changes from the year 2000 to the year 2010. Specifically, we use the global frustration, $f^{(G)}(n)$, and vexation, $v^{(C)}$, parameters from the year 2000 (Fig.~\ref{fig:4}a), and the constraints, $\mu^{(C)}$, matched to county composition changes (Fig.~\ref{fig:4}b) over a given number of iterations defined by $\tau^{(C)}$ (Fig.~\ref{fig:4}c) to define, $H'(n)$ (Fig.~\ref{fig:4}d) used in Eqs.~\ref{eq:rates1},\ref{eq:rates2}. We then perform the appropriate number of iterations to forecast the probability of a neighborhood (Fig.~\ref{fig:4}e) with an initial composition in the year 2000 (Fig.~\ref{fig:4}f black line) to have any composition in the year 2010 (Fig.~\ref{fig:4}F blue line). These historic forecasts of the changes from 2000 to 2010 can then be compared to the observed changes (Fig.~\ref{fig:4}g) to investigate the accuracy of this method.

\begin{figure}
    \centering
    \includegraphics[width = 5.9in]{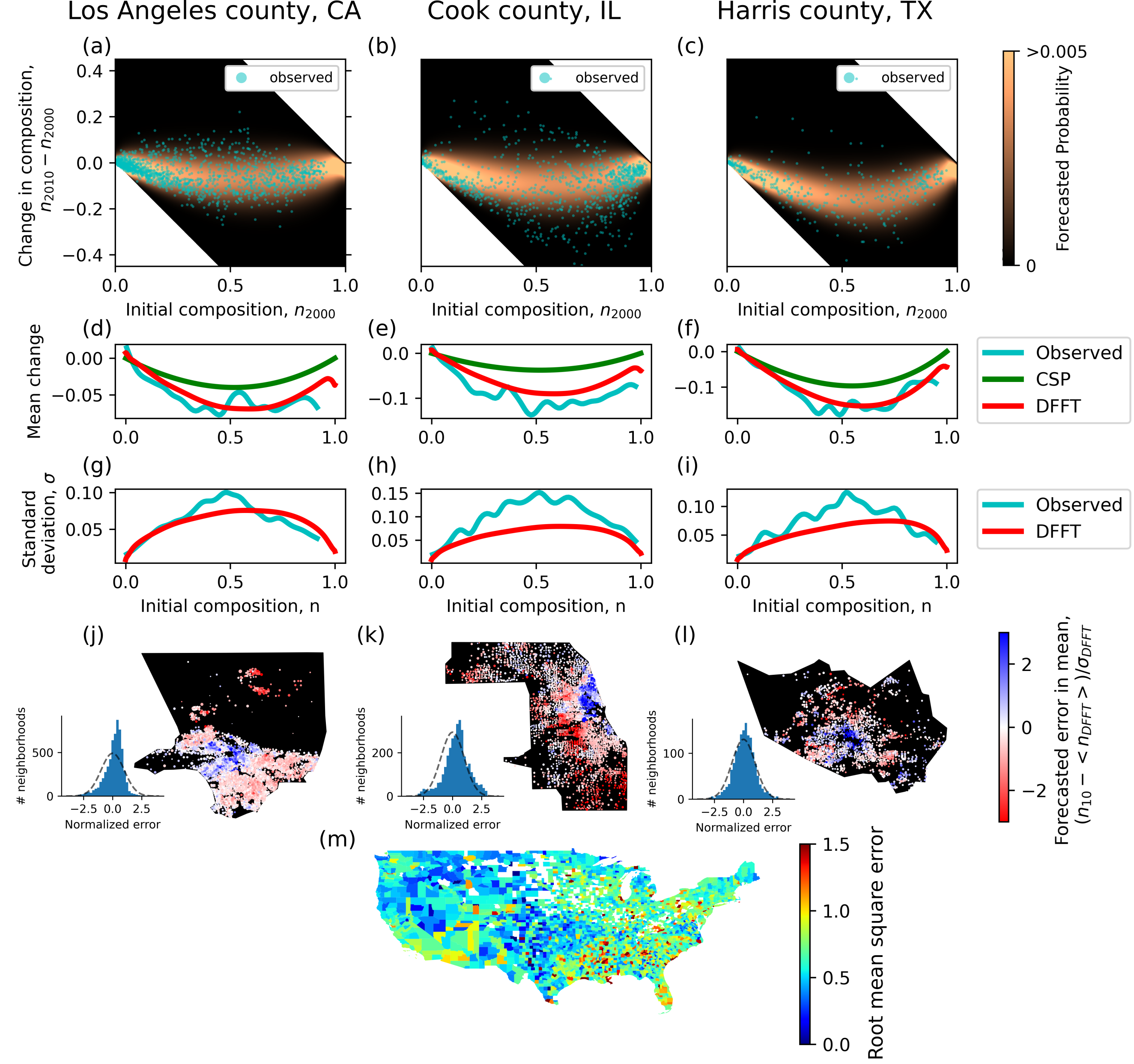}
    \caption{\textbf{Validation of DFFT forecasting method on 2010 census data for White/non-White racial grouping.}
    (\textbf{a-c}) Forecasts (heatmaps) of the probability of observing changes in neighborhood composition from 2000 to 2010 (y-axis) as a function of initial composition (x-axis) for the three most populous counties (Los Angeles, Cook, and Harris) using average neighborhood sizes for each county. Brighter shading indicates more probable change in composition. Forecasts are compared to observed changes in composition (scatter points) for neighborhoods within 20\% of the average neighborhood size. (\textbf{d-f}) Expected mean forecasted changes (red curves) plotted alongside weighted moving averages of observed changes (blue curves) and a standard small area population projection method, Constant Share of Population (green curves). DFFT forecasts agree with observed changes and improve upon the prior method. (\textbf{g-i}) Forecasted standard deviations (red lines) plotted alongside weighted moving standard deviations (blue lines). High values indicate greater tendencies for change. (\textbf{j-l}) Bubble charts geographically depicting differences between forecasted mean changes and observed changes normalized by the standard deviation forecasted by the model, $\sigma_{DFFT}$. Sizes of bubbles are proportional to neighborhood size $s$. (insets) Histograms of normalized differences within each county (blue bars) and theoretical gaussian distributions (gray dotted curve). Neighborhoods with similar errors are spatially clustered together at sub-county scales but approach gaussian distributions on county scales. (\textbf{m}) Root mean square errors gathered across each county. Counties with less than 2\% of either the White or non-White subgroup were excluded due to limited possible composition changes. 
    }
    \label{fig:5}
\end{figure}

We find our predictions for 2010 neighborhood composition changes based on 2000 neighborhood-scale data to be remarkably accurate. We illustrate this accuracy by presenting results for the 2000-2010 neighborhood composition changes of the White population for the three most populous counties in the US (LA county, CA, Cook county, IL, and Harris county, TX), which contain sufficient neighborhood samples to see trends in dynamics across nearly all compositions. The DFFT forecasts of the probability of compositional changes, shown in Fig.~\ref{fig:5}a-c (heatmaps), predict that integrated neighborhoods are both more susceptible to change (greater variance) and, on average, experience the largest decrease when compared to their segregated counterparts. These predictions arise from a combination of the tendency of segregated neighborhoods to resist changes to their composition with the simultaneous decrease in county composition of the White subgroup. Comparison of these forecasts to actual changes in composition (Fig.~\ref{fig:5}a-c blue scatter points) show that the same trends are actually present in the 2010 census data. To further demonstrate this quantitative agreement, we plot (Fig.~\ref{fig:5}d-f) the weighted moving averages of observed neighborhood changes (blue curves) along side the DFFT forecasted average change (red curves) and the change predicted from a standard method for small area population projections, Constant Share of Population (green lines)\cite{wilson2015new}. We find that the DFFT forecasts match the average changes in composition with impressive fidelity while the Constant Share of Population method is significantly less accurate. Beyond providing merely predictions of average neighborhood change, however, the DFFT framework can forecast the full probability distribution of observing any future composition. Importantly, these forecasted distributions provide additional information regarding the likelihood of any particular change in composition, thereby accounting for the impact at small populations sizes ($\sim$1000 individuals) of the randomness of birth, death, and migration events. Prediction of such probability distributions are typically not accessible in current small-area population projections such as the Constant Share of Population method. To visualize the likelihood of a neighborhood's change away from the mean, we plot the standard deviations (Fig.~\ref{fig:5}g-i) of the probabilistic DFFT forecasts (red lines) and the observed neighborhood changes (blue lines) and find good agreement. The close agreement indicates that our method can forecast both average changes in composition and the likelihood of observing any particular change in neighborhood composition. 

To determine whether this framework introduces unexpected spatial biases both within counties and across coarser regions in the US, we compare the spatial distribution of the prediction errors both at the neighborhood and county scales. To allow for comparisons between neighborhood changes independently of their initial compositions, we calculate differences between the observed changes (Fig.~\ref{fig:5}a-c scatter points) and DFFT forecasted mean changes (Fig.~\ref{fig:5}d-f red curves), divide these residuals by the standard deviations forecasted by DFFT (Fig.~\ref{fig:5}g-i) and plot these normalized errors geospatially as bubbles (Fig.~\ref{fig:5}j-l) whose size is proportional to the number of persons in a neighborhood. Spatial correlations in the normalized errors at sub-county scales represent shared properties between nearby neighborhoods that the DFFT framework does not account for. However, we find the distribution of these normalized errors collected across the county (Fig.~\ref{fig:5}j-l (insets)) approach Gaussian distributions with a mean of 0 and a standard deviation of 1 suggesting that DFFT are reliable measures of neighborhood change. Lastly, we take the root mean square of these normalized errors over each county and compare them over the entire nation. We find these root mean square errors to be roughly consistent across the United States ($0.87 \pm 0.06$ when weighted by population) indicating the accuracy of our forecasts extend beyond the three counties shown here.

\begin{figure}
    \centering
    \includegraphics[width = 6in]{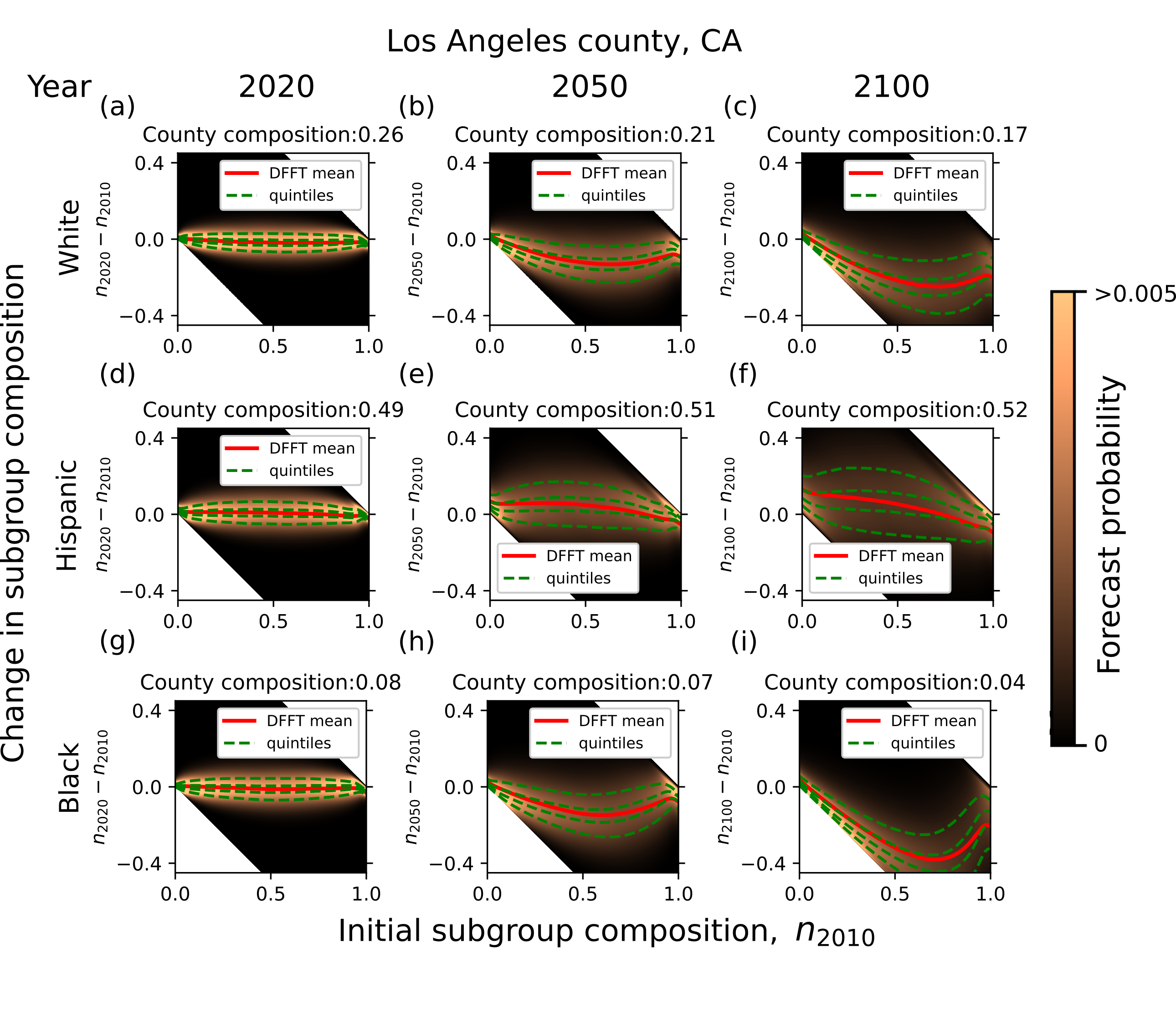}
    \caption{\textbf{Forecasts of unknown composition changes.}
    DFFT forecasts of the same type as Fig.5~(a-c) into the future for White (\textbf{a-c}), Hispanic (\textbf{d-f}), and Black (\textbf{g-i}) subgroups in Los Angeles county, CA for the years 2020 (\textbf{a,d,g}), 2050 (\textbf{b,e,h}), and 2100 (\textbf{c,h,i}) using projected changes in county composition\cite{hauer2019population}. Red curves track the DFFT forecasted mean and green dotted curves divide vertical shaded probability distributions into equal probability quintiles. As time evolves, forecasts become more spread. These forecasts do not account for errors in population projections nor potential changes in segregation but, instead, only account for accumulated statistical randomness from neighborhood changes over time.
    }
    \label{fig:6_races}
\end{figure}

Encouraged by these results we turn our framework towards forecasting the neighborhood compositions for the years 2020 and beyond. Importantly, the analysis performed here was conducted prior to the release of the 2020 census results and therefore serves as an additional blind test of the DFFT approach. Since we do not know the changes in county compositions ahead of time, we employ traditional population projections developed at the county scale \cite{hauer2019population} to approximate the expected change in county composition in time. We then use exactly the same methodology described previously (Fig.~\ref{fig:4}) except now using 2000 and 2010 data to forecast neighborhood compositions changes for 2020. Separate example forecasts for White, Hispanic, and Black populations in LA county are shown in Figure~\ref{fig:6_races}. 
For the White subgroup (Fig.~\ref{fig:6_races}a), 
we present DFFT forecasts of the probability of compositional changes (heatmaps) from 2010 to 2020. Similar to the forecast of 2010 (Fig.~\ref{fig:5}a), due to the persistence of segregation, integrated neighborhoods are expected to experience greater changes than segregated neighborhoods, as shown by the increased vertical spread of the heatmaps and quantified by the accompanying quintiles (green dotted lines). In addition to the 2020 projections, it is instructive to project such forecasts further into the future to portray the ultimate impact that current segeregation drivers could have on our population if they remain. Towards this end, we show forecasts for the years 2050 (Fig.~\ref{fig:6_races}b) and 2100 (Fig.~\ref{fig:6_races}c) constrained to match projected county compositions for each intervening decade. As expected, we find that forecasted probability distributions broaden over time with increasing spread between quintiles. Similar trends are observed for the Hispanic and Black population subgroups (Fig.~\ref{fig:6_races}d-i). The most notable difference is that the means of the projections (red lines) for the Hispanic subgroup have an overall downward curvature due to an increase in county composition. Collectively, these results indicate that if current segregation drivers remain unchanged, neighborhoods with compositions at the extremes will not change significantly. In short, segregated neighborhoods will persist. 

\section{Discussion and Conclusion}
The success of the above forecasts at the neighborhood scale confirm that a segregation function, as opposed to an index, can yield new insights and predictions of residential demography. Moreover, we have shown that the probability distributions that result from these functions offer intuitive representations of segregation behavior (Fig.~\ref{fig:3}c,f,i, and the supplemental).

In this work, we measured segregation tendencies using a global frustration function that represents the dominant trends in segregation on the national scale. However, it is well known that segregation varies across counties. Through the use of multiple frustration functions to account for these differences, it may be possible to group together counties according to their frustration functions and generate more accurate forecasts. Such capabilities will provide a powerful new lens through which to study residential demographics.

Moreover, this framework easily lends itself to further extensions that can improve the accuracy of these small-area forecasts. For example, thus far we have divided populations into only one subgroup at a time (e.g. White and non-White). However, through the use of a multicomponent DFFT approach (\cite{ChenArxiv}), we could divide populations into multiple groups simultaneously (e.g. Black, White, and Hispanic). Such groupings have the potential to measure more detailed segregation behaviors that better reflect population diversity and further improve forecasts. Additionally, this framework relied solely on population counts. It is expected that understandings of social processes and forecasts could be dramatically improved if this framework included additional sources of data. For example, through study of the profiles of neighborhoods that exhibit similar DFFT errors, we can understand how neighborhood properties correlate with atypical changes. These correlations could then be incorporated in a model that generates forecasts based on both the current composition of a neighborhood and properties such as median income or housing type. Finally, by casting this problem of small-area projections into a statistical framework with well-defined measures for forecast accuracy, we open the door to further quantitative methods for improving predictions ranging from traditional demographic projections to machine learning. Such techniques could improve upon the already impressive accuracy obtained with this direct DFFT approach.   

Importantly, if forecasts shown here prove to be robust into the future, this framework can aid in the development of policies addressing persistent racial inequities. For example, forecasts of average changes in neighborhood composition (Fig.~\ref{fig:1}c) can inform decisions on the location of affordable housing, the allocation of school resources, investments in local infrastructures, and access to goods and services, with the aim of better meeting the needs of the future population. Furthermore, the vertical spread of probabilistic forecasts (Fig.~\ref{fig:5}a-c) helps identify neighborhoods most susceptible to potentially negative disruptions in their communities such as those associated with gentrification and residential tipping. This identification can inform the implementation of targeted policies such as housing assistance to ameliorate displacement. With this end in mind, we have made our forecasts available to any interested parties. Specifically, we have published online the forecasted probability distribution for each individual neighborhood and each racial/ethnic sub-grouping as well as an interactive open source geospatial app to explore forecasting results for any neighborhood within the entire United States. We anticipate that such data will enable better planning and allow for more streamlined policy implementation at the neighborhood scale. 

We anticipate that this novel approach using population counts to quantify segregation and forecast small-area projections may serve as an effective direction for policy analysis. For example, through county comparisons, one may feasibly correlate historical policies (e.g., redlining) with measurements of the resulting frustration functions. Such an analysis could go beyond simple index-based measurements and elucidate the impact of policies in greater detail. Moreover, this quantification could model how such decisions may affect the spatial distributions of population subgroups if implemented. More broadly, we hope that as the tools presented here are improved and developed further by the wider community, they will spur development of policies aimed at mitigating some of the more pernicious and persistent inequities that continue to plague the nation.

\section{Computational methods}
\subsubsection{Schelling simulation}
To generate the statistics shown in figure \ref{fig:2}, we employ a simulation similar to the one described in \cite{ChenArxiv}, but without spatial effects and with moves in which agents switch locations rather than move to empty cells. Briefly, the simulation behaves according to the following rules. First, a pair of agents of different types are chosen. Second, the number of red and blue 8-connected neighbors for each site are counted. Third, the total change in Utility, $\Delta U$, for the proposed move is calculated where $U(N)=0.3N$ is a linear function of the number of neighbors, $N$ of the same type. Fourth, the agents switch if the move increases their Utility $\Delta U>0$, or, if the move decreases their Utility or keeps it the same, $\Delta U \leq 0$, the agents switch with a probability $e^{-\Delta U}$. For more details and the code used to perform this analysis, see Code Availability section.

\subsubsection{Computational optimization of DFFT vexation and frustration parameters}
As it is rare to observe neighborhoods with the same total population, $s$, we use the sample size invariant penalty function shown in Eq.~\ref{eq:probability_distribution}, $H(n)$, which is the primary focus of our computational methods. More precisely, given a collection of counties, the penalty function is county specific, $H^{(C)}(n)$, where we optimize for county specific linear terms (vexations), $v^{(C)}$, and a global frustration, $f(n)$, which captures higher-order terms.  Given that the frustration is expected to be a continuous function of the composition, $n$, it can be discretized and presumed constant over sufficiently small compositional domains, $f(n) \rightarrow f_{D_n}$. Using maximum likelihood estimation over the vexations and discretized frustrations, 
\begin{equation}
    {\mathcal L}=\max_{f_{D_n}, v^{(C)}} \sum_{C}\sum_{b} \log\Big(P(n_b;s_b,f_{D_n},v^{(C)}) \Big),
    \label{eq:optimization}
\end{equation}
they can be determined through non-linear optimization given a set of observations. Here the outer sum is over all counties $C$ and the inner sum is over all census block groups $b\in C$ within the county. $n_b$ and $s_b$ represent the observed composition and neighborhood size, respectively, and $f_{D_n}$ is the appropriate discretized frustration term given the observation falls within its domain boundaries, $n_b \in D_n$, and $P(n_b;s_b,f_{D_n},v^{(C)})$ is the corresponding probability. The binning for the discretization need not have constant spacing: in practice, it is prudent to have smaller domains near compositional extremes due to fact that the compositions of many census block groups are dominated by one ethnic/racial subgroup. Once a discretized frustration terms $f_{D_n}$ are obtained, they can be made continuous through an appropriate interpolation (e.g., cubic spline).  
\\
An efficient approach to speed up the optimization is to employ Sterling's approximation for the binomial coefficient,
\begin{equation}
    \binom{s}{sn} \approx e^{-s\Big(nln(n) + (1-n)ln(1-n) \Big)}
    \label{eq:Sterling}.
\end{equation}
The advantage of this approach is that it allows the entropy factor, $nln(n) + (1-n)ln(1-n)$, to be absorbed into $H^{(C)}(n)$. The sum of the entropy factor and the $H^{(C)}(n)$ function we refer to as $\overline{H}^{(C)}(n)$. The form of equation (\ref{eq:probability_distribution}) for a single county then becomes
\begin{equation}
    P(n;s) = z^{-1}\binom{s}{sn}e^{-sH(n)} \approx z^{-1}e^{-s\overline{H}(n)},\label{eq:probability_distribution2}
\end{equation}
where $z$ remains the appropriate normalization. Specifically, because county probability distributions are generally not sharply peaked, consistent with segregation, it is advisable to discretize $\overline{H}^{(C)}(n)$ and keep the relationship to vexation and frustration terms implicit. Differentiation of the likelihood with respect to $f_{D_n}$ leads to a maximum likelihood condition for each domain,
\begin{equation}
    \sum_C \sum_b s^{(C)}_b \delta(n^{(C)}_b \in D) = \sum_C \sum_s\frac{ s N^{(C)} (s)N_s(n \in D)e^{-s \overline{H}^{(C)}_D}}{\sum_{D'} N_s(n \in D') e^{-s \overline{H}^{(C)}_{D'}}},
\end{equation}
where the left side corresponds to the observed data and the right side to the behavior of the model. In this equation, $n^{(C)}_b$ and $s^{(C)}_b$ is the the observed composition and size of neighborhood \textit{b} in county \textit{C}. $N^{(C)}(s)$ is the number of times that a neighborhood of size \textit{s} occurs in the county, $N_s (n \in D)$ is the number of possible ways that a neighborhood of size \textit{s} can end up in the domain \textit{D} and $\delta(n^{(C)}_b \in D)$ is defined as 1 if $n^{(C)}_b$ is in the domain \textit{D} and 0 otherwise.

On the other hand, differentiation of the likelihood with respect to $v^{(C)}$ leads to a maximum likelihood condition for each county,
\begin{equation}
     \sum_b s^{(C)}_b n^{(C)}_b =  \sum_s\frac{ \sum_D \overline{n}^{(C)}_D s N^{(C)} (s)N_s(n \in D)e^{-s \overline{H}^{(C)}_D}}{\sum_D N_s(n \in D) e^{-s \overline{H}^{(C)}_D}},
\end{equation}

where $\overline{n}^{(C)}_D $ is the average observation in county \textit{C} and domain \textit{D}.

It is now possible to define an error function based on how far the parameters deviate from these conditions, find the gradients, and perform a more efficient non-linear optimization than would be possible using the exact solution, represented by Eq.\ref{eq:optimization}. 

\subsubsection{Detailed balance of dynamic DFFT forecasting model} 

Here we demonstrate that the recursive equations used for forecasting (Eqs.~\ref{eq:rates1},~\ref{eq:rates2}) obey detailed balance, thus guaranteeing that neighborhood probability distributions will eventually evolve to the distribution in equation \ref{eq:probability_distribution} under reasonable conditions. Detailed balance states that, at equilibrium, every possible process is in equilibrium with its reverse. For our model this gives
\begin{equation}
\begin{aligned}
    P(n+\Delta n)p(n+\Delta n\rightarrow n) = P(n)p(n\rightarrow n + \Delta n) \quad \forall n,
\end{aligned} \label{detblance}
\end{equation}
where $P(n)$ is the probability that the neighborhood has composition $n$, and  $p(n\rightarrow n + \Delta n)$ is the probability that a neighborhood with $n$ changes to $n + \Delta n$ within a given time-step. Because neighborhoods have discrete compositions which, provided a sufficiently small time-scale, change one move at a time, we can consider limit our considerations to one-person increments: $\Delta n  = 1/s \equiv \delta n$. The recursive equations are thus
\begin{align}
    p(n \rightarrow n - \delta n) &= \frac{n}{1+e^{-s \big(H(n) - H(n-\delta n) \big) }} \label{eq:rates11}\\
     p(n \rightarrow n + \delta n) &= \frac{1-n}{1+e^{s \big(H(n+ \delta n) - H(n) \big)}}, \label{eq:rates21}
\end{align}
where the derivatives of the recursive equations (\ref{eq:rates1},\ref{eq:rates2}) are represented by the appropriate finite difference. The detailed balance condition from \ref{detblance} can then be rewritten as
\begin{equation}
\begin{aligned}
    \frac{P(n+ \delta n)}{P(n)} = \frac{p(n \rightarrow n + \delta n)}{p(n + \delta n \rightarrow n )}.
\end{aligned} 
\end{equation}
The ratio of consecutive binomial coefficients on the left-hand side simplifies to $(1-n)/(n+\delta n)$, and the identity $(1+e^{-x})/(1+e^{x}) = e^{-x}$ produces $e^{-s\big(H(n+\delta n)- H(n) \big)}$ on the right-hand side, so that each side can be seen to be $\frac{(1-n)}{(n+\delta n)} e^{-s\big(H(n+\delta n)- H(n) \big)}$, thereby establishing detailed balance and demonstrating that the transition probabilities indeed lead to the expected probability distribution.

\subsubsection{Transition matrix implementation of recursive forecasting equations}
For convenience and computational efficiency, we incorporate the recursive equations (Eqs.~\ref{eq:rates1},~\ref{eq:rates2}) into a transition matrix, $T$, that evolves forecasted probability vectors into the future. Representing the full probability distribution of observing a given composition as a probability vector $\lambda = [P(n)]$, the transition matrix $T$ forecasts future probability distributions using simple matrix multiplication,
\begin{equation} \label{eq:transition_matrix}
    \lambda_{t+1} = T \lambda_t,
\end{equation}
from the initial condition
\begin{equation*}
        \lambda_0 = \begin{cases}
        1    &\text{if \(n=n_0\)}\\
        0    &\text{otherwise} \end{cases},
\end{equation*}
where the composition is known exactly. Eq.~\ref{eq:transition_matrix} then allows rapid prediction of the time evolution of the state vector $\lambda$ into the future.

\subsubsection{Optimization of the $\mu^{(C)}$ and $\tau^{(C)}$ parameters}
To match forecasts of the form generated using equation \ref{eq:transition_matrix} onto observed dynamics, we simultaneously account for shifts in county compositions (Fig.~\ref{fig:4}b) while aligning our dynamic DFFT framework onto the decade time scale (Fig.~\ref{fig:4}c). To constrain forecasts to a desired future county composition after a chosen number of iterations $i^{(C)}$ of equation~\ref{eq:transition_matrix}, we choose the appropriate value of the county composition constraint $\mu^{(C)}$. To find this value quickly we note that if the number of iterations are applied in proportion to the neighborhood size then the expected compositional change, to lowest order, is neighborhood-size independent. This allows us to obtain a good approximation of $\mu^{(C)}$ using only the average neighborhood size in the county. Then, using the initial probability vector, $\lambda^{(C)}$ and a given number of iterations, $i^{(C)}$, we use a binary search to find the value of $\mu^{(C)}$ that, when incorporated with the transition matrix, $T$, shifts the expected value of the probability vector from the initial county composition to the desired final county composition, e.g. $\langle n_{final}\rangle = \sum_n (T_{\mu}^i\lambda^{(C)})n$. As expected, when the proposed number of iterations, $i^{(C)}$, is small, then the absolute value of $\mu^{(C)}$ is large in order to quickly drive the county composition to the desired final composition. 

Finally, to determine the correct total number of iterations $i$ for evolving neighborhood compositions, we optimize a time constant $\tau^{(C)}$ (units of $ \# \text{ of iterations}/\text{person}/\text{year}$) using historical composition changes. For the underlying transition matrix $T$, a single iteration corresponds to the average rate at which a single individual move is considered. For example, the number of iterations needed for a neighborhood with $s$ persons over a 10 year period is $i=10 s \tau^{(C)}$, thus allowing for more individual moves for neighborhoods with greater populations. It is important to keep in mind that, beyond proportionality to population size, many local factors may influence the rate of moves, from housing type to local culture. To determine the optimal time constant, we consider a list of trial time constants from $0.1$ to $3$. We then calculate the corresponding values of $\mu^{(C)}$ for a given proposed county composition change and each trial time constant. Then, for each county, we take each individual neighborhood $b$ with size $s_b$ and initial composition $n_{b,0}$ and, from it, generate a forecast using Eq.~\ref{eq:transition_matrix} and the number of iterations defined by the trial time constant in order to forecast the full probability distribution $\lambda_{b}$. From these forecasted probability distributions, $\lambda_{b}$, for each time constant, we can extract the probability of the observed final composition, $n_{b,f}$. We assume neighborhood size stays the same and thus round the final composition to the nearest possible composition for the initial neighborhood size. Lastly, we sum the log of these probabilities over the neighborhoods within a county to build a log-likelihood metric, $\mathcal{L}^{(C)}$, of the forecast for each candidate time constant,
\begin{equation}
    \mathcal{L}^{(C)}=\sum_b log(\lambda_b(n_{f};\tau^{(C)})),
    \label{eq:lglik_score}
\end{equation}
where the sum is over all neighborhoods $b \in C$ in county $C$. We then choose the time constant that leads to the greatest log-likelihood value. 
The observed values for these time constants can vary from 0.1 to 2.5. Such variations are likely due to differences in mobility within different counties, (urban counties typically have greater mobilities and time constants than rural counties), differences in segregation between counties, and the effect of redistributions of subgroups within counties. Time constants for counties are shown in the Supplemental Information.

\subsubsection{Statistics of forecasting accuracy}
We develop here a summary statistic for the accuracy of our forecasts of the block group compositions in the year 2010 to open the door to future methods that may surpass results presented here. Since our forecasts are probabilistic, we use a log-likelihood metric,
\begin{equation}
    \mathcal{L}=\frac{\sum_b log(\lambda_b(n_{f}))}{\sum_b s_b},
    \label{eq:lglik_score_nation}
\end{equation}
where the sum is over neighborhoods $b$ from all counties in the US and $\lambda_b(n_{f})$ is the probability of observing the composition $n_{f}$ in the year 2010. Since the log-likelihood roughly scales with the population size, we normalize by the total population of all the included neighborhoods, $\sum_b s_b$, in the year 2000 to allow for more flexibility in making comparisons. Furthermore, to address outliers due to data quality issues as discussed in the Data Availability section, we apply a log-likelihood floor value of -10 and do not include neighborhoods whose total population changed by more than 25\%. We thus obtain values of $\mathcal{L} = 4.13008\times 10^{-3}, 3.73931\times 10^{-3}, \text{and } 3.41130 \times 10^{-3}$ for the White, Hispanic, and Black subgroupings respectively. These summary statistics for forecasts alongside the data used to generate them serve as a baseline for researchers across disciplines to improve upon.

\subsubsection{Uncertainties in frustration functions and hypothetical probability distributions}
To approximate uncertainties in frustration functions inferred from population counts, we subsample counties using bootstrapping methods. Specifically, we randomly select 100 subsets of counties that contain at least 50 million total people. Then, we extract frustration functions that best fit population counts from each of these subsets and use those functions and their resulting probability distributions to generate the shading region representing one standard deviation shown in figures~\ref{fig:3} and in Supplemental figures. While segregation certainly varies between counties, these uncertainties represent the confidence of the nationwide average frustration function, $f^{(G)}(n)$. 

\subsection{Code availability}
Code for analysis performed here is written in Python and/or Julia and can be found at \url{https://osf.io/tgxcr/?view_only=4826e86ce189420d846c8f16b934ac2c}. Within this repository are also open-source tools for users to geospatially explore both the accuracy of the forecasts of the 2010 census results and to generate forecasts into the future.
Furthermore, up-to-date code for the Schelling simulation analysis as used for figure \ref{fig:2} alongside examples are available at \url{https://github.com/yunuskink/DFFT-Schelling-model}.

\subsection{Data availability}
Raw data, processed data, and forecasts for each neighborhood can be found at \url{https://osf.io/tgxcr/?view_only=4826e86ce189420d846c8f16b934ac2c} including results for the non-contiguous parts of the US not shown in figures here.
Raw census block group population counts for 1990, 2000, and 2010 were obtained through NHGIS \cite{NHGIS}. This dataset was used to infer DFFT functions as shown in figure \ref{fig:3}. Furthermore, we use longitudinally tracked population counts approximately aligned to 2010 block group geometries developed by NHGIS\cite{NHGIS} to optimize the time scales, $\tau^{(C)}$, and also to validate the accuracy of the forecast as shown in figure \ref{fig:5}. However, such alignment of block group geometries over time are prone to significant errors\cite{LOGAN2021102476}. For example, dense populations such as correctional facilities and dormitories can be moved from one neighborhood to an adjacent neighborhood leading to unrealistic changes in residential compositions. We found neighborhoods whose total population changed by less than 25\% had no clear alignment error and so only include those neighborhoods for optimizing time scales. We additionally set a floor for the minimum log-probability in Eq.~\ref{eq:lglik_score} to $-10$ to reduce the weight given to potential outliers not filtered out by changes in total neighborhood population. For example, counties with zero Black persons in 2000 technically have infinite vexations, $v^{(C)} = \infty$. If Black persons moved into such a county, then unrealistically poor forecasts result, and setting such a minimum probability minimized the impact of such outlier events.

Projections of changes in county scale composition used in Figs.~\ref{fig:1},~\ref{fig:6_races} were calculated using the SSP2 forecast from Hauer (2019). 

\subsection{Author contributions}
Y.A.K conceived application to demographic data, performed analysis, wrote initial draft, and produced figures. Y.A.K. and B.B. developed and refined the theory. Y.A.K. and B.B. developed algorithms to perform statistical fitting and generate forecasts. I.C. mentored Y.A.K. and T.A.A. mentored B.B.. M.H. guided development of model onto census data and placed research into context of policy development. All authors contributed to final draft of text. 

\subsection{Acknowledgements}
The authors would like to acknowledge Yuchao Chen for preliminary analysis not shown here. Y.A.K. was supported in part by funding from the National Science Foundation Graduate Research Fellowship Award (DGE-1650441). I.C. and Y.A.K. were also supported in part by funding from (NINDS, 1R01NS116595) and Army Research Office (ARO W911NF-18-1-0032). B.B. was supported in part by the Natural Sciences and Engineering Research Council (NSERC PGS D). 

\bibliographystyle{unsrt}
\newpage
\setcounter{page}{0}
\ \ \ \ \ \ \ \ \ \ \ \ \ \ \ \ \ \ \ \ \ \ \ \ \ \ \ \ \ \ \ \ \ \ \ \ \ \ \ \ \ \ \ {\bf BIBLIOGRAPHY}
\bibliography{references}

\begin{thebibliography}{10}

\bibitem{wellman2014transportation}
Gerard~C Wellman.
\newblock Transportation apartheid: the role of transportation policy in
  societal inequality.
\newblock {\em Public Works Management \& Policy}, 19(4):334--339, 2014.

\bibitem{anderson2003providing}
Laurie~M Anderson, Joseph~St Charles, Mindy~T Fullilove, Susan~C Scrimshaw,
  Jonathan~E Fielding, Jacques Normand, Task~Force on~Community
  Preventive~Services, et~al.
\newblock Providing affordable family housing and reducing residential
  segregation by income: a systematic review.
\newblock {\em American journal of preventive medicine}, 24(3):47--67, 2003.

\bibitem{oka2014capturing}
Masayoshi Oka and David~WS Wong.
\newblock Capturing the two dimensions of residential segregation at the
  neighborhood level for health research.
\newblock {\em Frontiers in public health}, 2:118, 2014.

\bibitem{kershaw2011metropolitan}
Kiarri~N Kershaw, Ana~V Diez~Roux, Sarah~A Burgard, Lynda~D Lisabeth, Mahasin~S
  Mujahid, and Amy~J Schulz.
\newblock Metropolitan-level racial residential segregation and black-white
  disparities in hypertension.
\newblock {\em American journal of epidemiology}, 174(5):537--545, 2011.

\bibitem{massey1993american}
Douglas Massey and Nancy~A Denton.
\newblock {\em American apartheid: Segregation and the making of the
  underclass}.
\newblock Harvard university press, 1993.

\bibitem{krysan2017cycle}
Maria Krysan and Kyle Crowder.
\newblock {\em Cycle of segregation: Social processes and residential
  stratification}.
\newblock Russell Sage Foundation, 2017.

\bibitem{hauer2016millions}
Mathew~E Hauer, Jason~M Evans, and Deepak~R Mishra.
\newblock Millions projected to be at risk from sea-level rise in the
  continental united states.
\newblock {\em Nature Climate Change}, 6(7):691--695, 2016.

\bibitem{wilson2016evaluation}
Tom Wilson.
\newblock Evaluation of alternative cohort-component models for local area
  population forecasts.
\newblock {\em Population Research and Policy Review}, 35(2):241--261, 2016.

\bibitem{smith2013practitioner}
Stanley~K Smith, Jeff Tayman, and David~A Swanson.
\newblock {\em A practitioner's guide to state and local population
  projections}.
\newblock Springer, 2013.

\bibitem{wachter2014essential}
Kenneth~W Wachter.
\newblock {\em Essential demographic methods}.
\newblock Harvard University Press, 2014.

\bibitem{raftery2012bayesian}
Adrian~E Raftery, Nan Li, Hana {\v{S}}ev{\v{c}}{\'\i}kov{\'a}, Patrick Gerland,
  and Gerhard~K Heilig.
\newblock Bayesian probabilistic population projections for all countries.
\newblock {\em Proceedings of the National Academy of Sciences},
  109(35):13915--13921, 2012.

\bibitem{crone2011advances}
Sven~F Crone, Michele Hibon, and Konstantinos Nikolopoulos.
\newblock Advances in forecasting with neural networks? empirical evidence from
  the nn3 competition on time series prediction.
\newblock {\em International Journal of forecasting}, 27(3):635--660, 2011.

\bibitem{vasan2018use}
Srini Vasan, Jack Baker, and Ad{\'e}lamar Alc{\'a}ntara.
\newblock Use of kernel density and raster manipulation in gis to predict
  population in new mexico census tracts.
\newblock {\em Review of Economics \& Finance}, 14:25--38, 2018.

\bibitem{chi2017small}
Guangqing Chi and Donghui Wang.
\newblock Small-area population forecasting: a geographically weighted
  regression approach.
\newblock In {\em The frontiers of applied demography}, pages 449--471.
  Springer, 2017.

\bibitem{mendez2018density}
J~Felipe M{\'e}ndez-Valderrama, Yunus~A Kinkhabwala, Jeffrey Silver, Itai
  Cohen, and TA~Arias.
\newblock Density-functional fluctuation theory of crowds.
\newblock {\em Nature communications}, 9(1):3538, 2018.

\bibitem{ChenArxiv}
Yuchao Chen, Yunus~A Kinkhabwala, Boris Barron, Matthew Hall, Tomas~A Arias,
  and Itai Cohen.
\newblock Forecasting the dynamics of segregated population distributions at
  the neighborhood scale using density-functional fluctuation theory.
\newblock {\em arXiv preprint arXiv:2008.09663}, 2020.

\bibitem{ellen2016cityscape}
Ingrid Ellen and Lei Ding.
\newblock Cityscape: A journal of policy development and research: Symposium on
  gentrification.
\newblock {\em Cityscape}, 2016.

\bibitem{freeman2009neighbourhood}
Lance Freeman.
\newblock Neighbourhood diversity, metropolitan segregation and gentrification:
  What are the links in the us?
\newblock {\em Urban Studies}, 46(10):2079--2101, 2009.

\bibitem{crowder2000racial}
Kyle Crowder.
\newblock The racial context of white mobility: An individual-level assessment
  of the white flight hypothesis.
\newblock {\em Social Science Research}, 29(2):223--257, 2000.

\bibitem{poot2016gravity}
Jacques Poot, Omoniyi Alimi, Michael~P Cameron, and David~C Mar{\'e}.
\newblock The gravity model of migration: the successful comeback of an ageing
  superstar in regional science.
\newblock {\em IZA discussion paper}, 2016.

\bibitem{hauer2019population}
Mathew~E Hauer.
\newblock Population projections for us counties by age, sex, and race
  controlled to shared socioeconomic pathway.
\newblock {\em Scientific data}, 6(1):1--15, 2019.

\bibitem{grauwin2012dynamic}
Sebastian Grauwin, Florence Goffette-Nagot, and Pablo Jensen.
\newblock Dynamic models of residential segregation: An analytical solution.
\newblock {\em Journal of Public Economics}, 96(1-2):124--141, 2012.

\bibitem{armstrong2001extrapolation}
J~Scott Armstrong.
\newblock Extrapolation for time-series and cross-sectional data.
\newblock In {\em Principles of forecasting}, pages 217--243. Springer, 2001.

\bibitem{wilson2015new}
Tom Wilson.
\newblock New evaluations of simple models for small area population forecasts.
\newblock {\em Population, Space and Place}, 21(4):335--353, 2015.

\bibitem{tanton2014review}
Robert Tanton et~al.
\newblock A review of spatial microsimulation methods.
\newblock {\em International Journal of Microsimulation}, 7(1):4--25, 2014.

\bibitem{baker2014spatial}
Jack Baker, Ad{\'e}lamar Alc{\'a}ntara, Xiaomin Ruan, Kendra Watkins, and Srini
  Vasan.
\newblock Spatial weighting improves accuracy in small-area demographic
  forecasts of urban census tract populations.
\newblock {\em Journal of Population Research}, 31(4):345--359, 2014.

\bibitem{james1985measures}
David~R James and Karl~E Taeuber.
\newblock Measures of segregation.
\newblock {\em Sociological methodology}, 15:1--32, 1985.

\bibitem{massey1988}
Douglas~S. Massey and Nancy~A. Denton.
\newblock {The Dimensions of Residential Segregation*}.
\newblock {\em Social Forces}, 67(2):281--315, 12 1988.

\bibitem{reardon2002measures}
Sean~F Reardon and Glenn Firebaugh.
\newblock Measures of multigroup segregation.
\newblock {\em Sociological methodology}, 32(1):33--67, 2002.

\bibitem{reardon2004measures}
Sean~F Reardon and David O’Sullivan.
\newblock Measures of spatial segregation.
\newblock {\em Sociological methodology}, 34(1):121--162, 2004.

\bibitem{yao2019spatial}
Jing Yao, David~WS Wong, Nick Bailey, and Jonathan Minton.
\newblock Spatial segregation measures: A methodological review.
\newblock {\em Tijdschrift voor economische en sociale geografie},
  110(3):235--250, 2019.

\bibitem{s2016segregation}
Christopher S.~Fowler.
\newblock Segregation as a multiscalar phenomenon and its implications for
  neighborhood-scale research: The case of south seattle 1990--2010.
\newblock {\em Urban geography}, 37(1):1--25, 2016.

\bibitem{hennerdal2017multiscalar}
Pontus Hennerdal and Michael~Meinild Nielsen.
\newblock A multiscalar approach for identifying clusters and segregation
  patterns that avoids the modifiable areal unit problem.
\newblock {\em Annals of the American Association of Geographers},
  107(3):555--574, 2017.

\bibitem{chodrow2017structure}
Philip~S Chodrow.
\newblock Structure and information in spatial segregation.
\newblock {\em Proceedings of the National Academy of Sciences},
  114(44):11591--11596, 2017.

\bibitem{roberto2015divergence}
Elizabeth Roberto.
\newblock The divergence index: A decomposable measure of segregation and
  inequality.
\newblock {\em arXiv}, pages arXiv--1508, 2015.

\bibitem{ellis2018predicting}
Mark Ellis, Richard Wright, Lee Fiorio, and Steven Holloway.
\newblock Predicting neighborhood racial change in large us metropolitan areas,
  1990--2010.
\newblock {\em Environment and Planning B: Urban Analytics and City Science},
  45(6):1022--1037, 2018.

\bibitem{parisi2011multi}
Domenico Parisi, Daniel~T Lichter, and Michael~C Taquino.
\newblock Multi-scale residential segregation: Black exceptionalism and
  america's changing color line.
\newblock {\em Social Forces}, 89(3):829--852, 2011.

\bibitem{NHGIS}
Steven manson, jonathan schroeder, david van riper, tracy kugler, and steven
  ruggles. ipums national historical geographic information system: Version
  15.0 [dataset]. minneapolis, mn: Ipums. 2020.
\newblock \url{http://doi.org/10.18128/D050.V15.0}.

\bibitem{LOGAN2021102476}
John~R. Logan, Wenquan Zhang, Brian~J. Stults, and Todd Gardner.
\newblock Improving estimates of neighborhood change with constant tract
  boundaries.
\newblock {\em Applied Geography}, 132:102476, 2021.

\end{thebibliography}

\newpage

\renewcommand{\theequation}{S\arabic{equation}}
\renewcommand{\thefigure}{S\arabic{figure}}
\renewcommand{\thetable}{S\arabic{table}}
\renewcommand{\thesection}{S\arabic{section}}

\setcounter{equation}{0}
\setcounter{figure}{0}
\setcounter{table}{0}
\setcounter{section}{0}




\begin{figure}
    \centering
    \includegraphics[width = 5in]{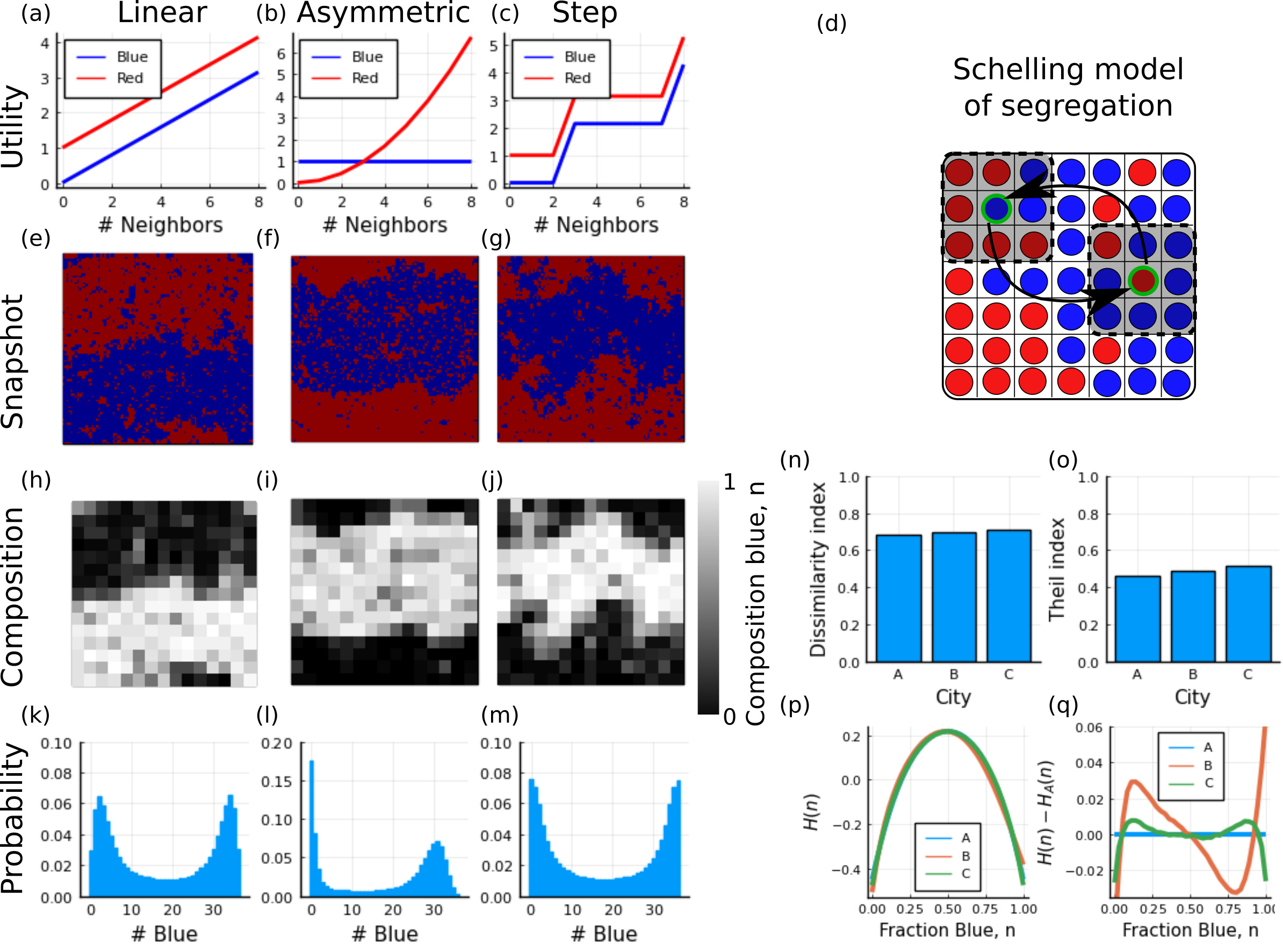}
    \caption{\textbf{Comparison between traditional segregation indices and DFFT functions.}
    (\textbf{a-c}) Utility functions defined for Schelling segregation model. Utilities for a given agent are calculated based on the number of 8-connected neighbors that are the same type as the chosen agent. Increasing functions drive segregation. (\textbf{d}) Underlying Schelling model. Red and blue agents choose to switch locations based on utility changes incurred by proposed switches. (\textbf{e-g}) Sample snapshots from simulations with equal number of red and blue agents using utility functions (\textbf{a-c}), respectively. (\textbf{h-j}) Neighborhood composition of blue agents from (\textbf{e-g}), respectively, when grouped into 6x6 size neighborhoods. (\textbf{k-m}) Respective probability distributions of neighborhood compositions tallied over 1000 snapshots and all neighborhoods. (\textbf{n-o}) Dissimilarity and Theil segregation indices\cite{reardon2002measures} measuring the degree of segregation between blue and red agents. (\textbf{p}) $H$ functions obtained through transformation of probability distributions shown in (k-m) using Eq.~1 in the main text. (\textbf{q}) $H$ functions shown in (p) with the "linear" $H$ function subtracted. Probability distributions (k-m) reveal that each city has a different behavior due to differences in the utility functions (a-c) despite having nearly identical segregation indices (n-o). In contrast, the DFFT $H$ functions (p-q) retain all the information presented in the probability distributions and can be used to forecast dynamic changes in simulated neighborhood compositions\cite{ChenArxiv}.
    }
    \label{fig:si_seg_index}
\end{figure}

\begin{figure}
    \centering
    \includegraphics[width = 5in]{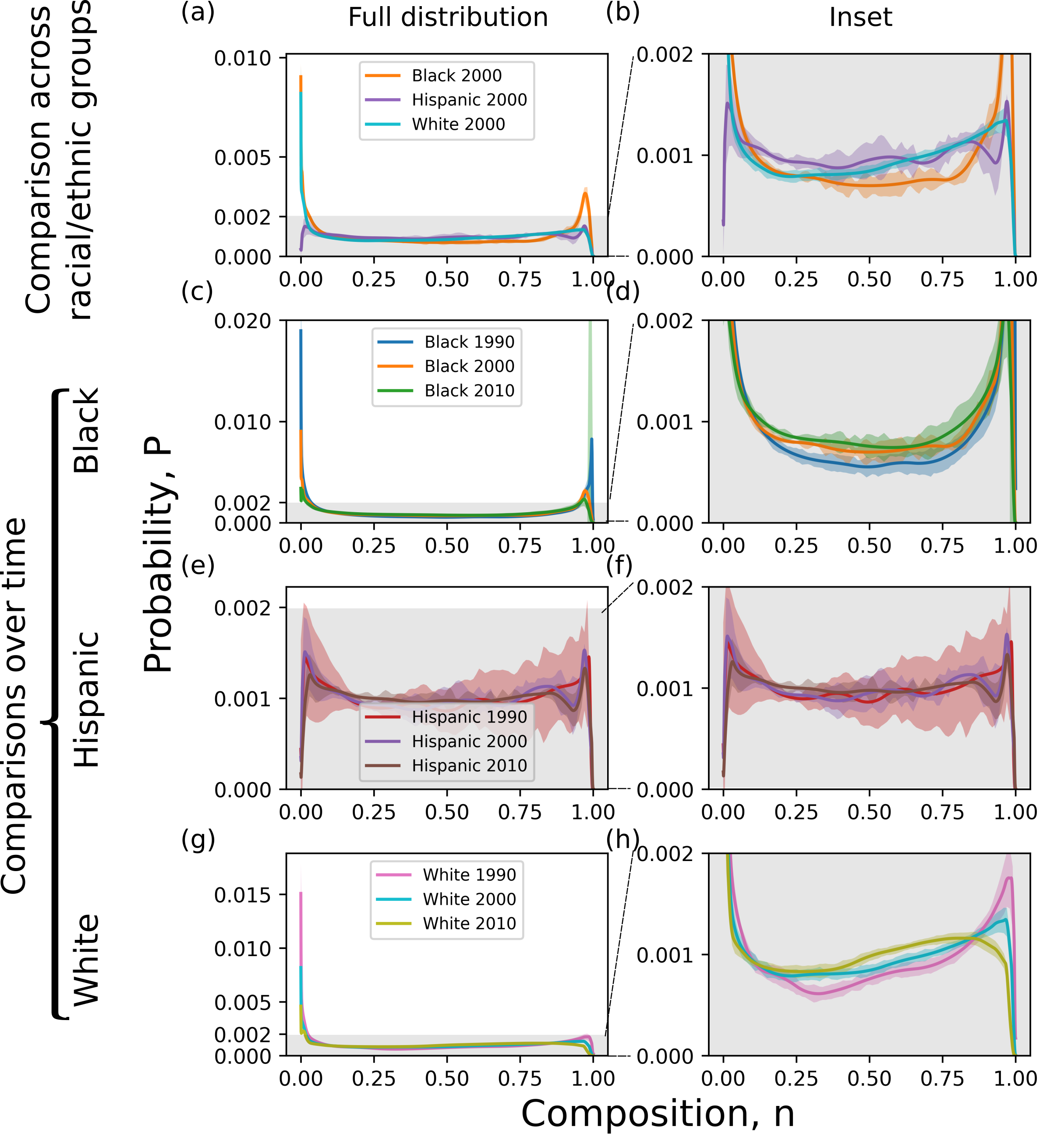}
    \caption{\textbf{Frustration function reveals national trends in segregation both between different racial/ethnic groups and over time.}
    (\textbf{left column}) Probability distributions inferred using global frustration functions, $f^{(G)}(n)$, and vexation values tuned to match county compositions of 0.5 of the appropriate ethnic subgroup along with zoomed in insets (\textbf{right column}) to better visualize differences between distributions. (\textbf{a,b}) Comparison between Black/non-Black, Hispanic/non-Hispanic, and White/non-White distributions from 2000 census. The Hispanic grouping has the greatest probability for integrated compositions while Black subgroup has the least. Additionally, the White subgroup distribution increases from minority White to majority White capturing the tendency of neighborhoods towards either majority White compositions or predominantly non-White compositions near zero. (\textbf{c,d}) Black/non-Black segregation for years 1990, 2000, and 2010. Probabilities of integrated compositions increased and segregated compositions decreased from 1990 to 2010. (\textbf{e,f}) Hispanic/non-Hispanic segregation for years 1990, 2000, and 2010. Segregation remained relatively constant from 1990-2010, possibly driven by a balance between decreasing segregation between long-time residents and increased immigration of Hispanic persons. (\textbf{g,h}) White/non-White segregation for years 1990, 2000, and 2010. Probabilities of predominantly White compositions near one decreased and probabilities of integrated neighborhoods from 0.25-0.75 increased from 1990-2010. Uncertainties represent one standard deviation calculated using bootstrapping methods.
    }
    \label{fig:si_comparison}
\end{figure}

\begin{figure}
    \centering
    \includegraphics[width = 5in]{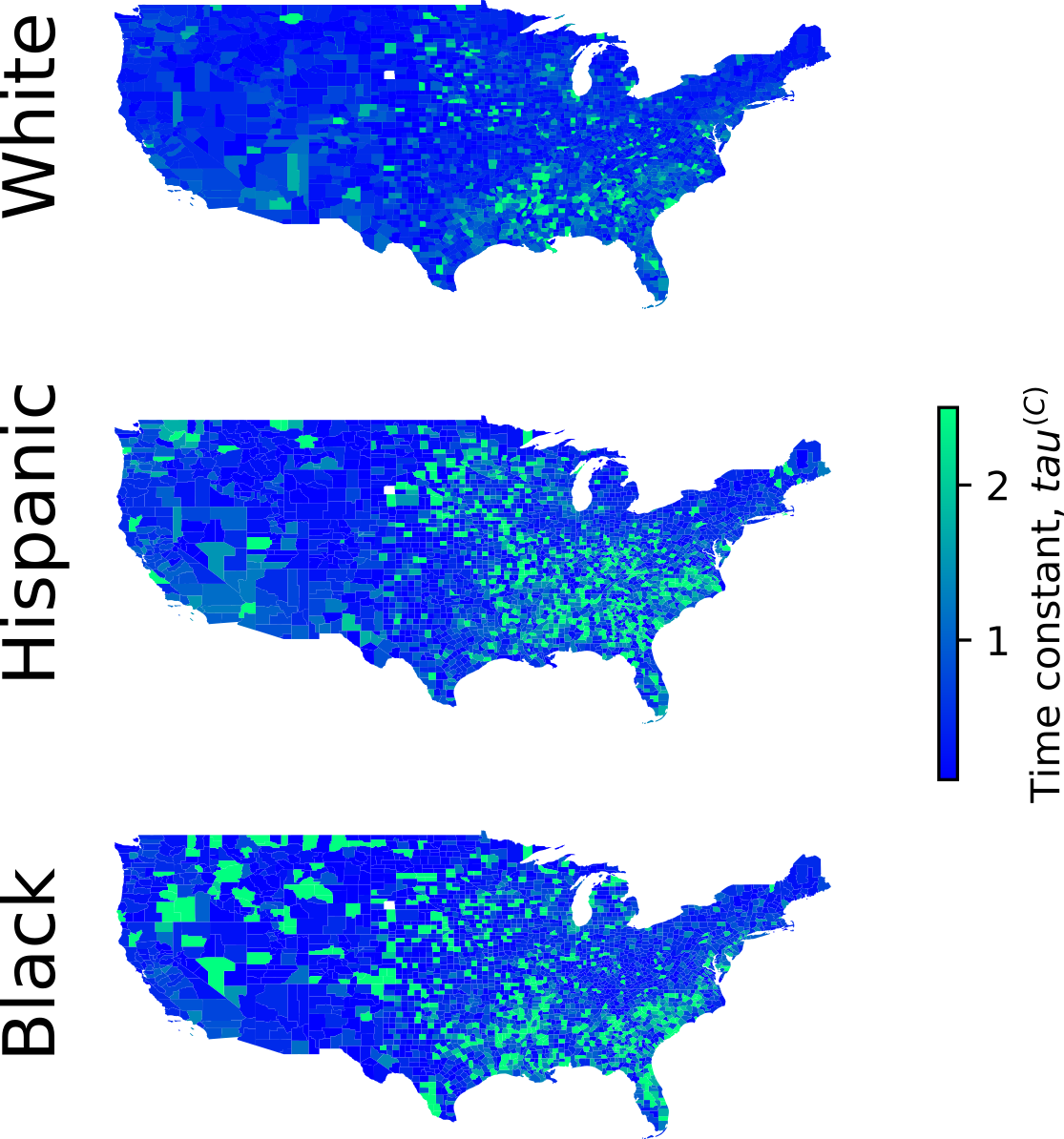}
    \caption{\textbf{Time constants, $\tau^{(C)}$.}
    County time constants $\tau^{(C)}$ (choropleth) resulting from analysis of composition changes from 1990 to 2000 for each of the three racial/ethnic subgroups, White, Hispanic, and Black. Counties dominated by a single population subgroup have very small time constants due to insufficient data regarding dynamic changes. For example, in the extreme case where all neighborhoods are populated entirely by one subgroup in both the year 1990 and 2000, then the best time constant is zero, guaranteeing that the compositions of neighborhoods do not change. 
    }
    \label{fig:si_taus}
\end{figure}

\begin{figure}
    \centering
    \includegraphics[width = 5in]{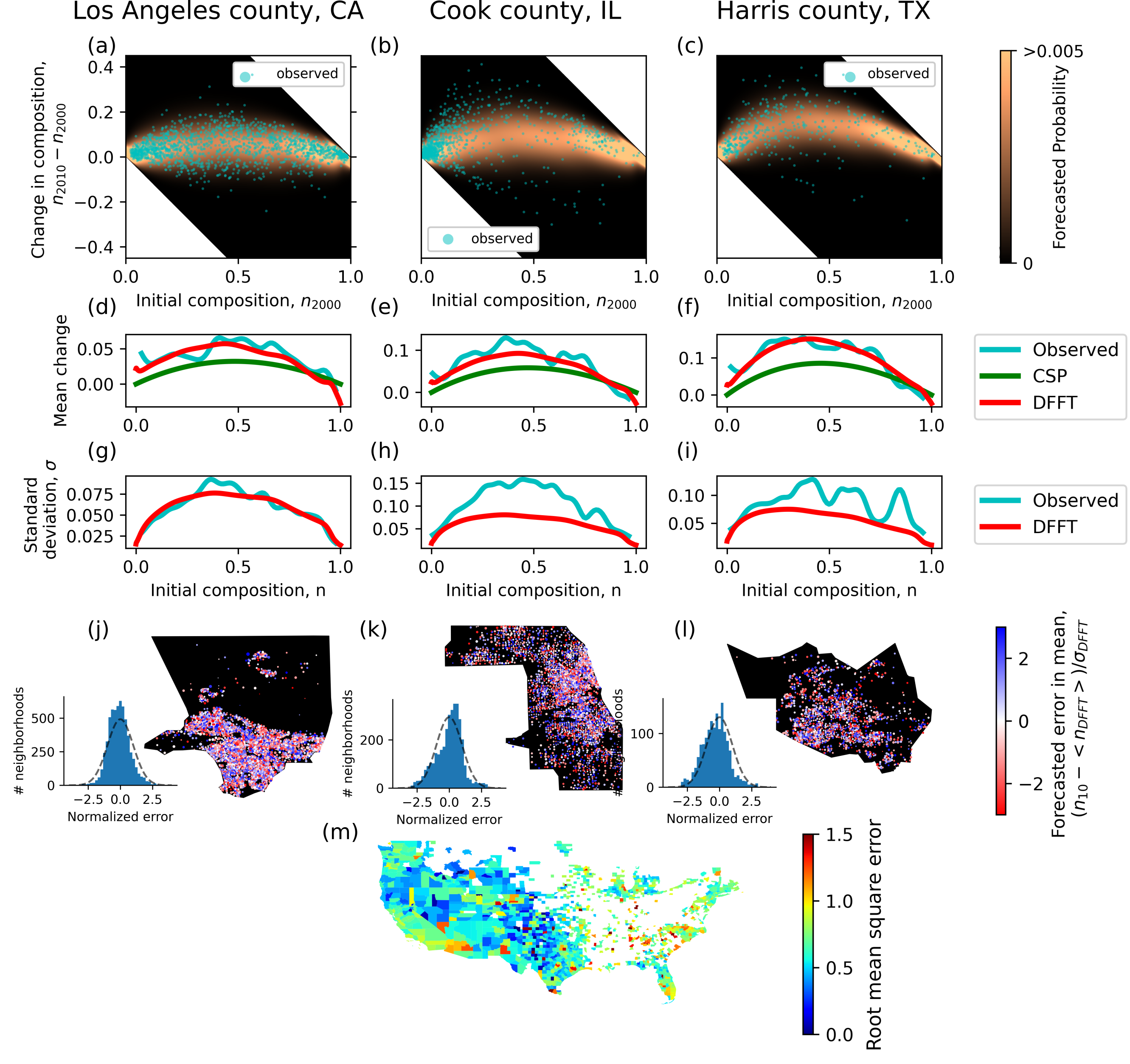}
    \caption{\textbf{Validation of DFFT forecasting method on 2010 census data for Hispanic/non-Hispanic racial grouping.}
   Corresponds to Fig.~\ref{fig:5} from the main text but for the Hispanic subgroup. Due to increases in Hispanic county compositions from 2000-2010, we observe average increases in neighborhood compositions as opposed to the decreases for the White subgroup seen in the main text. 
    }
    \label{fig:si_hispanic}
\end{figure}

\begin{figure}
    \centering
    \includegraphics[width = 5in]{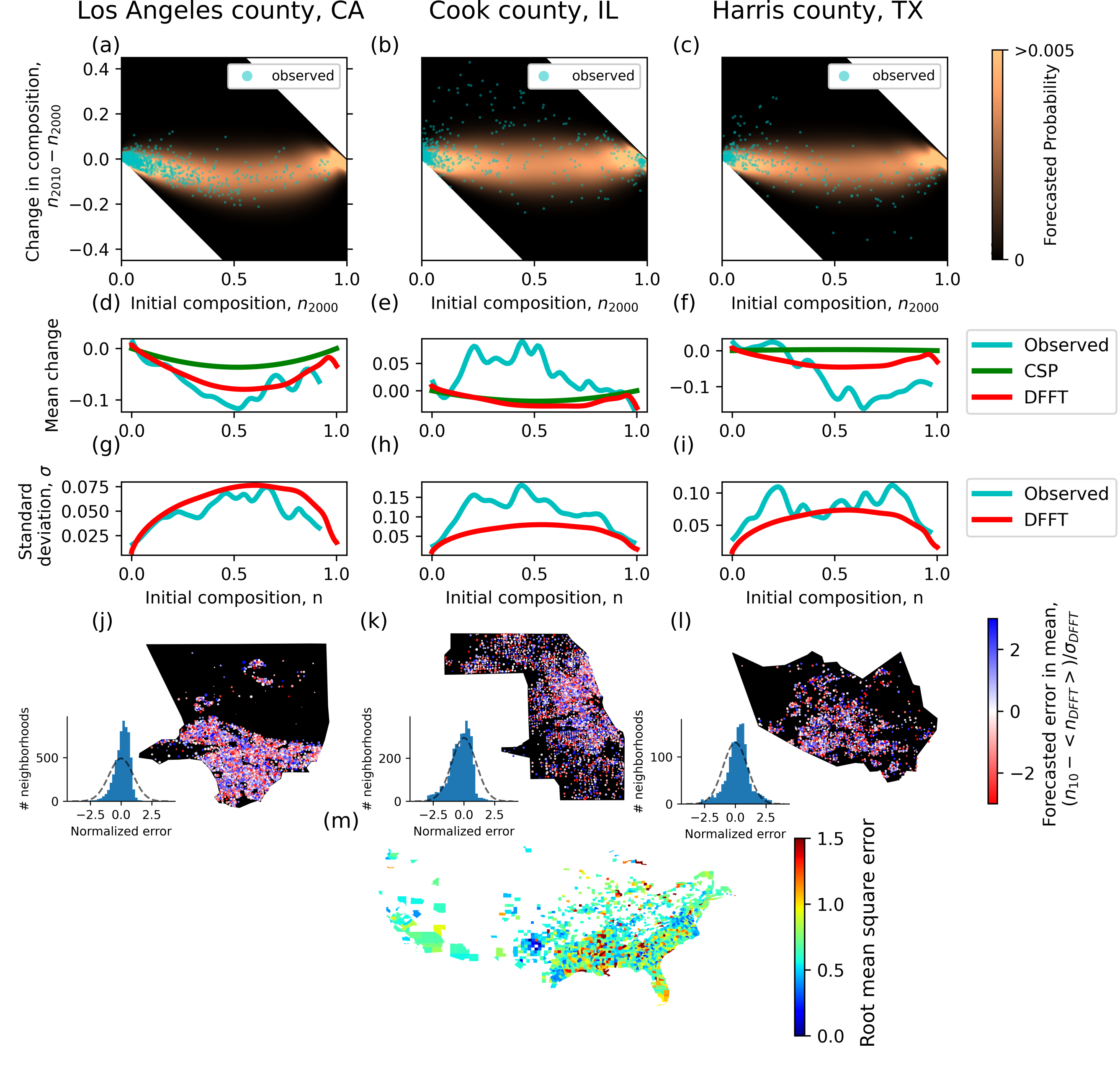}
    \caption{\textbf{Validation of DFFT forecasting method on 2010 census data for Black/non-Black racial grouping.}
    Corresponds to Fig.~\ref{fig:5} from the main text but for the Black subgroup. The apparent poor agreement between the mean change forecasts in (\textbf{e,f)} is due, in part, to the sparsity of data for intermediate compositions evident in the low number of data points at these compositions in (\textbf{b,c)}. 
    }
    \label{fig:si_black}
\end{figure}

\end{document}